\documentclass[a4paper,unpublished,onecolumn,11pt]{quantumarticle}
\pdfoutput=1
\usepackage[utf8]{inputenc}
\usepackage[english]{babel}
\usepackage[T1]{fontenc}
\usepackage{amsmath}
\usepackage{amssymb}
\usepackage{physics}

\usepackage{hyperref}
\usepackage{colortbl}
\usepackage[table]{xcolor}
\usepackage{booktabs}
\usepackage{enumerate}
\usepackage{enumitem}
\usepackage{caption}
\usepackage[normalem]{ulem}

\usepackage{amsthm}
\usepackage{caption}
\usepackage{selinput}
\SelectInputMappings{
  adieresis={ä}, % etc.
  acute={´},
}
\usepackage{amsmath}  % if not already loaded

\usepackage[english]{babel}
\usepackage[T1]{fontenc}
\usepackage{lipsum}
\usepackage{mdframed}
\usepackage{booktabs}
\usepackage{adjustbox}
\usepackage{longtable}
\usepackage{tikz}
\usetikzlibrary{arrows.meta,positioning,calc,fit,backgrounds}
\usepackage{pgfplots}
\pgfplotsset{compat=1.17}

\usetikzlibrary{arrows.meta,positioning,calc,fit,backgrounds}
\usepackage{pgfplots}
\pgfplotsset{compat=1.17}
\usepackage[numbers,sort&compress]{natbib}
\usepackage{hyperref}

\usepackage{pgfplots}
\pgfplotsset{compat=1.17}
\usepgfplotslibrary{groupplots}
\usepackage[table]{xcolor}
\usepackage{listings}
\usepackage{siunitx}
\usepackage{multicol}
\usepackage{pifont}  
\usepackage{newunicodechar}
\newunicodechar{✖}{\ding{55}}
\usepackage[margin=1in]{geometry}   % Adjust page margins
\usepackage{setspace}              % For line spacing
\usepackage{bm}
\usepackage{amsmath,amssymb,amsthm}             % Theorem and proof environments
\usepackage{tikz}            % TikZ is required by quanti
\usepackage{quantikz} 
\usepackage{enumitem}% For drawing quantum circuits
\usepackage{graphicx} 
\usepackage{algpseudocode}
\usepackage{algorithm}
\usepackage{algorithmicx}
\usepackage{xcolor}
\usepackage[most]{tcolorbox}
\usepackage{pgfplots}
\pgfplotsset{compat=1.17}

\usepackage{booktabs}

\usepackage{float}                 % For [H] float placement
\usepackage{hyperref}              % Clickable links/references
\usepackage{url}
\usepackage[nameinlink,noabbrev]{cleveref}

\usepackage{xcolor}                % For colored text (optional)
\usepackage{lscape}                % For landscape pages if needed
\usepackage{caption}
\usepackage{subcaption}
\usepackage{etoolbox}
\AtBeginEnvironment{abstract}{\raggedright}
\newtheorem{theorem}{Theorem}

\newtheorem{proposition}[theorem]{Proposition}
\newtheorem{corollary}[theorem]{Corollary}
\newtheorem{definition}[theorem]{Definition}

\theoremstyle{remark}

\newcommand{\OH}{\mathcal{H}_{\mathrm{OH}}}
\DeclareUnicodeCharacter{0301}{}
\makeatletter
\g@addto@macro\normalsize{%
    \abovedisplayskip 3pt plus 1pt minus 1pt%
    \abovedisplayshortskip 3pt plus 1pt minus 1pt%
    \belowdisplayskip 3pt plus 1pt minus 1pt%
    \belowdisplayshortskip 3pt plus 1pt minus 1pt%
}
\makeatother

\def\BibTeX{{\rm B\kern-.05em{\sc i\kern-.025em b}\kern-.08em
    T\kern-.1667em\lower.7ex\hbox{E}\kern-.125emX}}

\author{Chinonso Onah}
\affiliation{Volkswagen AG, Berliner Ring 2, 38440 Wolfsburg, Germany}
\affiliation{Department of Physics, RWTH Aachen University, 52056 Aachen, Germany}

\author{Stuart Hadfield}
\affiliation{USRA Research Institute for Advanced Computer Science (RIACS), CA, USA}

\author{Kristel Michielsen}
\affiliation{Forschungszentrum Jülich, Germany}
\affiliation{Universit\"at zu K\"oln, 50923 K\"oln, Germany}

\title{Leaving Quantum Bias on the Table: Pushforwards as a Path to Quantum Advantage in Constrained Optimization}

\begin{document}

\maketitle

\begin{abstract}
Constrained quantum samplers often produce infeasible outputs
that retain information about good feasible solutions.
We study the distribution obtained when a fixed classical
repair rule is applied to every measured sample and develop
a systematic post-measurement framework in which deterministic
classical repair pushes the raw quantum sampling law forward
onto the feasible set. Within the Onah--Firt--Michielsen
kernel, constraint statistics provide a framework to construct
problem-dependent repairs. We formulate these repairs as
retractions onto the feasible set and show that their
pushforward action defines a repair-induced sampling law
whose probability at each feasible configuration is the
total raw quantum mass of its repair preimage. We derive exact identities for repaired success and expected
cost. For noisy implementations, we separate encoded-subspace
support from leakage and track the degradation of
concentration on useful repair preimages, thereby
characterizing the loss of quantum bias and the approach
to repair-induced random sampling. For
permutation-constrained optimization problems, we specialize
the framework to Hungarian maximum-overlap repair and apply the formalism to the
traveling-salesman problem (TSP). We show that combining
repair-basin aggregation with amplitude transport and
constructive interference yields repaired-success lower
bounds, and derive an enhanced logarithmic-depth threshold
for repair-basin compensation. Our simulations show that optimal repair basins contain
selectively overweighted infeasible contributors. 
For an $n=10$ TSP instance, the optimum was
recovered using approximately $10^3$ shots
with pushforward repair, compared with the
$10^6$ shots used in the original CE--QAOA
workflow without repair. On a 127-qubit IBM processor,
using problem encodings of up to 100 qubits, repaired
candidates match the reference tours on the
smaller benchmarks and improve them on the larger instances.
These results establish concentration of
quantum probability on useful repair preimages as a
design principle for constrained quantum optimization.
\end{abstract}

\section{Introduction}
\label{sec:introduction}
Constrained quantum optimization is often implemented as a hybrid pipeline in which a quantum circuit produces candidate strings followed by classical objective evaluation. However, in highly constrained applications such as routing, assignment, scheduling, matching, and related combinatorial problems with hard structural constraints, the rate of direct feasible samples can remain small even when the sampled strings exhibit strong instance-dependent structure. To exploit this structure, we propose that problem-dependent classical processing of measured samples forms a third layer of the hybrid algorithm, following mixer transport and phase-sensitive interference~\cite{onahseperating}. This layer can expose constraint information present in the measured distribution, guide feasibility-restoring operations, and separate the contribution of the quantum sampler from that of subsequent classical processing. Such a layer is especially relevant for samples produced on near-term devices, where noise can further reduce the feasible-hit rate while leaving a large population of biased, noisy, or infeasible outcomes~\cite{Maciejewski_2025,hadfield2026ndar,Maciejewski2026,Onah2026GeometryInformed}.

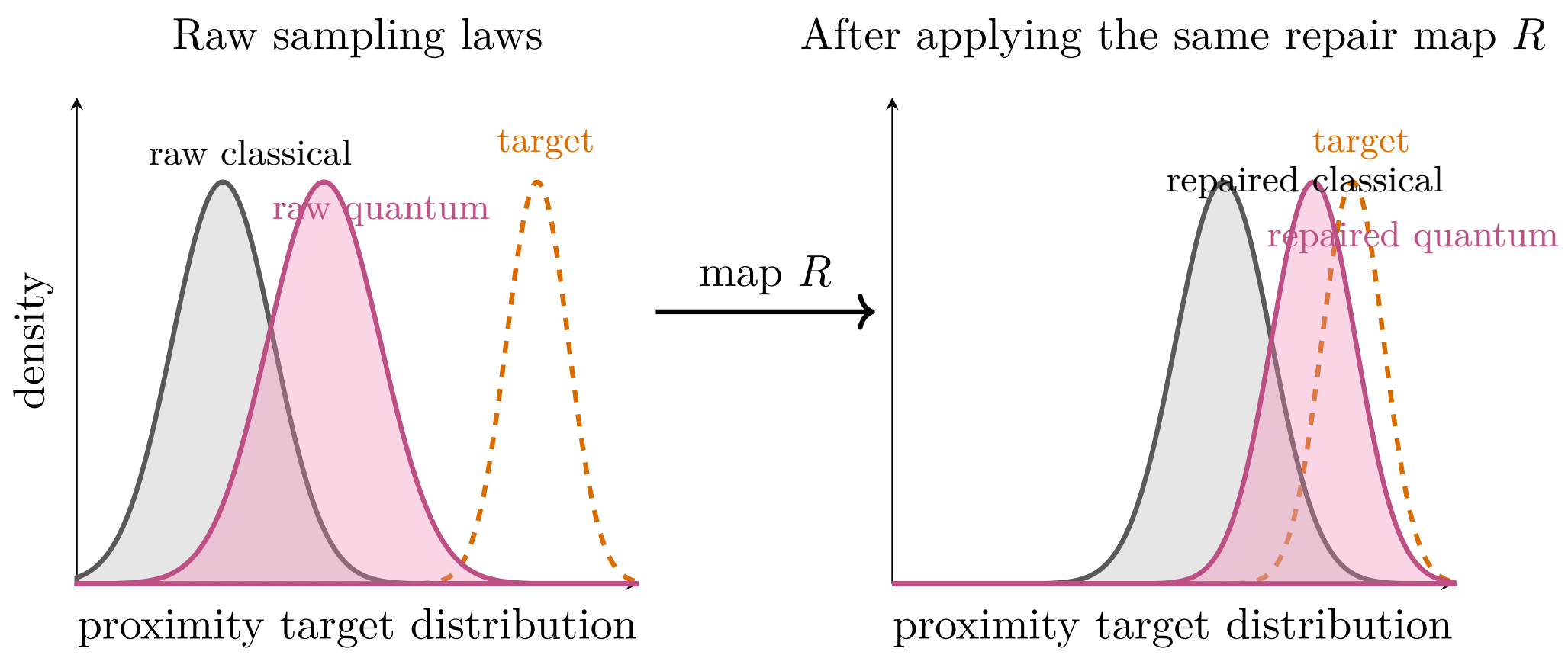
\begin{figure}[t]
\centering
\begin{tikzpicture}
\begin{groupplot}[
    group style={group size=2 by 1, horizontal sep=2.2cm},
    width=0.43\textwidth,
    height=5.8cm,
    xmin=0, xmax=10,
    ymin=0, ymax=1.15,
    axis lines=left,
    ytick=\empty,
    xtick=\empty,
    enlargelimits=false,
    clip=false
]

% ---------------- Left panel: raw laws ----------------
\nextgroupplot[
    title={Raw sampling laws},
    xlabel={proximity target distribution},
    ylabel={density}
]

% target
\addplot[
    domain=0:10,
    samples=220,
    very thick,
    dashed,
    draw=orange!85!black
]
{0.95*exp(-((x-8.2)^2)/(2*0.55^2))};

% classical raw
\addplot[
    domain=0:10,
    samples=220,
    very thick,
    draw=gray!70!black,
    fill=gray!55,
    fill opacity=0.35
]
{0.95*exp(-((x-2.6)^2)/(2*0.90^2))} \closedcycle;

% quantum raw
\addplot[
    domain=0:10,
    samples=220,
    very thick,
    draw=magenta!75!black,
    fill=magenta!60,
    fill opacity=0.35
]
{0.95*exp(-((x-4.4)^2)/(2*1.00^2))} \closedcycle;

\node[anchor=west] at (axis cs:1.0,1.02) {\footnotesize raw classical};
\node[anchor=west,text=magenta!75!black] at (axis cs:3.2,0.88) {\footnotesize raw quantum};
\node[anchor=west,text=orange!85!black] at (axis cs:7.2,1.04) {\footnotesize target};

% ---------------- Right panel: repaired laws ----------------
\nextgroupplot[
    title={After applying the same repair map \(R\)},
    xlabel={proximity target distribution}
]

% target
\addplot[
    domain=0:10,
    samples=220,
    very thick,
    dashed,
    draw=orange!85!black
]
{0.95*exp(-((x-8.2)^2)/(2*0.55^2))};

% repaired classical
\addplot[
    domain=0:10,
    samples=220,
    very thick,
    draw=gray!70!black,
    fill=gray!55,
    fill opacity=0.35
]
{0.95*exp(-((x-5.9)^2)/(2*0.85^2))} \closedcycle;

% repaired quantum
\addplot[
    domain=0:10,
    samples=220,
    very thick,
    draw=magenta!75!black,
    fill=magenta!60,
    fill opacity=0.35
]
{0.95*exp(-((x-7.5)^2)/(2*0.75^2))} \closedcycle;

\node[anchor=west] at (axis cs:4.6,0.95) {\footnotesize repaired classical};
\node[anchor=west,text=magenta!75!black] at (axis cs:6.4,0.82) {\footnotesize repaired quantum};
\node[anchor=west,text=orange!85!black] at (axis cs:7.2,1.04) {\footnotesize target};

\end{groupplot}

\draw[->, very thick]
($(group c1r1.east)+(0.15,0.25)$)
--
node[above] {\(\text{map }R\)}
($(group c2r1.west)+(-0.15,0.25)$);

\end{tikzpicture}
\caption{Conceptual distributional picture. The target law is shown as the dashed orange dome. At the raw level, neither the classical nor the quantum sampler reproduces the target distribution exactly, but the quantum law is more strongly biased toward it. Applying the same deterministic repair map \(R\) to both samplers shifts both distributions toward the target. The repaired quantum law moves substantially closer, while the repaired classical law improves but remains farther from the target.}
\label{fig:pushforward-dome-picture}
\end{figure}

The natural question then is what probability distribution does this three-layer hybrid procedure actually produce? Despite the ubiquity of such workflows, this central probabilistic object is usually left implicit. Hybrid performance is often described informally in terms of the raw quantum sampling distribution together with a repair heuristic, as though repair were merely a secondary implementation detail. We argue that once a suitable deterministic repair map is applied to every sample, however, the operational output is the induced law on the feasible set obtained by transporting raw quantum mass through that map. In other words, the relevant output distribution is the pushforward of the raw quantum sampling law through deterministic repair, as illustrated schematically in Fig.~\ref{fig:pushforward-dome-picture} and numerically in
Sec.~\ref{subsec:noiseless-analysis}. We show that combining structured quantum dynamics with deterministic feasibility restoration can yield efficiency gains that neither the quantum sampler nor the classical repair procedure produces alone. In benchmark studies, using the same repair rule for quantum and reference samples makes their respective mass on good-solution preimages directly comparable. This permits both set-level comparisons through
repaired cost CDFs and solution-level comparisons through the repaired
probabilities assigned to individual feasible outputs which we exhibit in detail in Sec.~\ref{subsec:noiseless-analysis}.

Recent works have established the algorithmic value of
classical processing after quantum sampling.
Caha, Kliesch, and Koenig incorporated greedy
post-processing into the QAOA objective and analyzed
approximation and depth tradeoffs for MaxCut
~\cite{Caha2022TwistedHybrid}.
Shirai and Togawa incorporated constraint-restoring
post-processing into variational schedules, with
simulations and experiments on constrained problems
~\cite{Shirai2024Postprocessing}.
For the traveling-salesperson problem, Bourreau,
Fleury, and Lacomme developed an indirect representation
with problem-dependent classical decoding
~\cite{bourreau2025warmstartq,bourreau20232}. 
Bucher \emph{et al.} combined warm-started $XY$
mixers, iterative sampling, and greedy repair of
noisy hardware outputs~\cite{bucher2026iterative}.
Previous work by two of us incorporated polynomial-time
feasibility repair and objective scoring into CE--QAOA,
establishing finite-shot and noise-dependent guarantees
for hybrid constrained optimization
~\cite{Onah2026GeometryInformed}. 
Sachdeva \emph{et al.} investigated the distributional
effects of greedy classical improvement in large-scale
variational quantum optimization, including comparisons
with post-processed random sampling
~\cite{sachdeva2026integrated}.

The present work characterizes hybrid
quantum--classical optimization through the sampling
law induced on the feasible set by systematically
constructed, problem-dependent classical repair maps.
In the process, we enlarge the optimal-solution
sampling event from the feasible optimum set to its
complete repair preimage.
Analytically, we separate repair-basin geometry
from sampler-dependent probability concentration,
quantify the role of deterministic tie-breaking,
and establish a uniform stochastic-repair
reference for permutation constraints.
Combining repair-basin aggregation with the
amplitude-transfer formalism yields conditional
finite-depth and finite-shot success guarantees.
We further quantify reductions in sampling
requirements and circuit depth thresholds
arising from the enlarged success event.
Noiseless TSP simulations and superconducting
hardware measurements illustrate the resulting
concentration of probability on useful repair
preimages.

% \sh{[Think we need one more short paragraph explaining exactly what this paper contributes beyond earlier repair-aware methods. Should identify:
% - the separation of repair geometry from sampler concentration;
% - the role of deterministic tie-breaking and the exact stochastic reference;
% - the conditional transport/phase certificate;
% - the repaired-law case study.]}

\subsection{Basic consequences of the pushforward formulation}

The pushforward viewpoint is elementary, but it changes what should be counted as
algorithmic success. Once every measured configuration is repaired, the raw sample
space is partitioned into repair basins
\[
B_y=R^{-1}(\{y\}),
\qquad y\in F,
\]
and the probability of obtaining a feasible solution \(y\) is the total raw probability mass contained in \(B_y\). Thus, deterministic repair replaces the pointwise task of sampling \(y\) directly by the less restrictive task of sampling any raw configuration that repairs to \(y\). A large basin provides a large combinatorial capture region, while the quantum circuit determines how strongly that region is actually populated. 

Several basic properties make this interpretation especially attractive for near-term quantum sampling tasks. The pushforward \(R_\#p\) remains a normalized probability law, and the expected cost after repair is obtained simply by repairing each raw sample and then evaluating the cost of the resulting feasible solution. Since the repair map sends every sample into the feasible set and fixes feasible points, it is idempotent. Applying the same deterministic repair map to two raw sampling laws cannot
increase their total-variation distance. Repair therefore cannot manufacture a
distinction between a quantum sampler and its baseline, while changing the
optimization-relevant event from a feasible target set \(G\) to its larger raw
preimage \(R^{-1}(G)\), thereby exposing bias already present in the raw
sampling law. Finally, when \(R\) is polynomial-time computable, samplewise
repair adds only \(O(S T_R(n))\) classical work for \(S\) shots and therefore
preserves polynomial end-to-end runtime. These structural and computational properties are collected formally in Proposition~\ref{prop:pushforward-properties} of Subsection~\ref{app:exact-pushforward-properties}.

At the dynamical level, enlarging the successful event reduces the amount of coherent shell-transfer amplification that must be generated by the quantum circuit at a fixed shot and confidence budget. The resulting reduction in required circuit depth is particularly valuable because phase alignment becomes progressively more demanding as additional layers introduce a larger number of interfering path contributions. The classical repair map supplies combinatorial basin volume, whereas the quantum circuit supplies nonuniform concentration within good basins.

\paragraph{Paper roadmap.}
The remainder of the paper is organized as follows.
Section~\ref{subsec:ofm-primitives} introduces the nested OFM geometry,
identifies the multidimensional constraint statistics retained by the penalty
Hamiltonian, and develops problem-dependent feasibility repair as a classical
OFM primitive. Section~\ref{subsec:definition-repair-basins} formulates the
repaired output as a pushforward law, derives the exact basin and
repaired-success identities, and separates encoded support from leakage under
noise. Section~\ref{sec:case-study-tsp} specializes the framework to the TSP
using Hungarian maximum-overlap repair and derives the repaired-success bound
and logarithmic-depth threshold. Section~\ref{sec:numerical-analysis} studies
the repaired law in noiseless simulation and on noisy superconducting
hardware. Sections~\ref{subsec:discussion-pointwise-to-basin}--%
\ref{subsec:beyond-hybrid-design} discuss basin amplification,
phase-control requirements, the separation between combinatorial volume and
quantum concentration, and the implications for hybrid algorithm design.
Appendix~\ref{app:exact-pushforward-properties} collects the structural
properties of deterministic pushforwards, while
App.~\ref{appendix:tie-breaking-uniqueness}--%
\ref{app:problem-stat} address deterministic tie-breaking, full-cube
sampling, and problem-dependent repair statistics beyond the permutation
setting.

\section{Maximizing Problem Structure via OFM Primitives}
\label{subsec:ofm-primitives}

\subsection{Layers of Problem-Algorithm Codesign}

Many constrained combinatorial optimization problems possess a nested geometry,
\[
\mathcal F
\subseteq
\mathcal X
\subseteq
\Omega,
\]
where \(\Omega\) is the ambient binary space, \(\mathcal X\) is a structured
encoded configuration space, and \(\mathcal F\) is the  feasible set. For
one-hot product encodings, we write
\[
\mathcal X \cong [n]^m,
\]
where \(n\) is the number of local one-hot choices available in each register, and \(m\) is the number of registers, even though the same problem may be represented algebraically in an ambient
space containing \(mn\) qubits. The local cardinality constraints defining
\(\mathcal X\) and the residual global constraints defining
\(\mathcal F\subseteq\mathcal X\) play different algorithmic roles and should therefore be treated accordingly. 

The Onah--Firt--Michielsen (OFM) kernel introduced in Ref.~\cite{onahce} provides a framework for exploiting this nested structure throughout a hybrid quantum--classical workflow. It specifies the encoded subspace, its constraint symmetries, the compatible mixer family, and the structured decomposition of the diagonal Hamiltonian.  Crucially, local fixed-cardinality constraints are built into the initial state and preserved by the mixing operator. The remaining global feasibility conditions are retained as diagonal penalty terms and used \emph{dynamically} to generate problem-dependent phases. In this work, we study how the same constraint geometry can be used \emph{classically} after measurement to transport the raw sampling law onto the feasible set. We recall the formal definition next.

\begin{definition}[Onah--Firt--Michielsen kernel {\cite{onahce}}]
\label{def:kernel-requirement}
An optimization instance belongs to the
\emph{Onah--Firt--Michielsen kernel} if there exist
\(n,m\in\mathbb N\) and a one-hot encoder
\(\mathsf E_{\mathrm{1hot}}\) such that the quantum dynamics is initialized
in the fixed-Hamming-weight space
\[
\OH=(\mathcal H_1)^{\otimes m},
\qquad
\mathcal H_1
=
\operatorname{span}
\{\ket{e_1},\ldots,\ket{e_n}\},
\]
with one excitation in each block. The computational-basis states of
\(\OH\) are identified with words
\[
x=(x_1,\ldots,x_m)\in[n]^m,
\qquad
\ket{x}
=
\ket{e_{x_1}}\otimes\cdots\otimes\ket{e_{x_m}}.
\]
We denote this encoded outcome space by
\[
\mathcal X=[n]^m.
\]

In the encoded basis, the diagonal problem Hamiltonian decomposes as
\[
H_C
=
H_{\mathrm{pen}}
+
H_{\mathrm{obj}}
+
H'_{\mathrm{pen}},
\]
where \(H_{\mathrm{obj}}\) encodes the objective,
\(H_{\mathrm{pen}}\) contains the symmetry-preserving global constraint
penalties, and \(H'_{\mathrm{pen}}\) contains additional diagonal penalties
that need not preserve the full kernel symmetry. The following requirements
hold:
\begin{enumerate}[label=\textup{(\alph*)},leftmargin=2.2em]

\item \emph{Penalty structure.}
The Hamiltonian \(H_{\mathrm{pen}}\) is a sum of squared affine residual
one-hot, degree, or capacity penalties, optionally together with linear
forbids, with integer coefficients bounded by \(\operatorname{poly}(n)\).
Consequently,
\[
\operatorname{spec}(H_{\mathrm{pen}})
\subseteq
\{0,1,\ldots,t_{\max}\},
\qquad
t_{\max}=\operatorname{poly}(n).
\]

\item \emph{Pattern symmetry.}
The symmetry-preserving part \(H_{\mathrm{pen}}\) is invariant under block
permutations \(S_m\) and global symbol relabelings \(S_n\). It therefore
organizes the encoded space into structured penalty levels
\[
L_t
=
\left\{
x\in\mathcal X:
\langle x|H_{\mathrm{pen}}|x\rangle=t
\right\}.
\]
The term \(H'_{\mathrm{pen}}\) collects problem-dependent penalty
contributions that break some or all of these symmetries.

\item \emph{Mixer and initial state.}
The canonical one-hot mixer is the normalized block-local \(XY\)
Hamiltonian
\begin{equation}
\label{eq:mixer}
\widetilde H_{XY}^{(b)}
=
\frac{1}{n-1}
\sum_{1\leq u<v\leq n}
\left(
X_u^{(b)}X_v^{(b)}
+
Y_u^{(b)}Y_v^{(b)}
\right),
\end{equation}
which preserves \(\mathcal H_1\) and satisfies
\[
\bigl\|
\widetilde H_{XY}^{(b)}
\bigr\|
=
O(1)
\]
on the one-excitation sector. The corresponding initial state is
\begin{equation}
\label{eq:initial-state}
\ket{s_0}
=
\ket{s_{\mathrm{blk}}}^{\otimes m},
\qquad
\ket{s_{\mathrm{blk}}}
=
\frac{1}{\sqrt n}
\sum_{r=1}^{n}\ket{e_r}.
\end{equation}
\end{enumerate}
\end{definition}

 Def. \ref{def:kernel-requirement} extends naturally to product \(k\)-hot subspaces when the mixer is replaced by a normalized number-conserving block-local Hamiltonian preserving \(\mathcal H_k\) \cite{B_rtschi_2019,yu2026efficient}, and covers several NP-Hard problems such as the TSP, QAP, CVRP, shared transportation problems, generalized assignment and multiple-knapsack problems, and \(k\)-dimensional matching for \(k\geq3\)\cite{onah2025waas,Lucas2014,GareyJohnson1979,Karp1972,onahce}.

\subsubsection{OFM Primitives}
We call an operation an \emph{OFM primitive} when it is derived from, and compatible with, the encoded subspace, the residual constraint structure, or the geometry of the feasible set fixed by Def. \ref{def:kernel-requirement}. The principal primitives are:
\begin{enumerate}[label=\textup{(\roman*)},leftmargin=2.4em]

\item \emph{Encoded initialization primitives}, which prepare states supported on $\OH$  and hence enforce the local cardinality constraints at initialization.

\item \emph{Amplitude transport primitives}, such as normalized \(XY\)-type mixers, which move probability amplitude within \(\mathcal X\) while enforcing the local cardinality constraints.

\item \emph{Problem-dependent phase operators.} Structure-resolving phase primitives, generated by
\(H_{\mathrm{pen}}\), \(H_{\mathrm{obj}}\), and
\(H'_{\mathrm{pen}}\), which imprint global feasibility, objective, and symmetry-breaking information on the encoded amplitudes; and

\item \emph{CE--QAOA ansatz.} CE--QAOA denotes the variational circuit built from a composition of OFM primitives (i)–(iii) above and belongs, conceptually, to the quantum alternating-operator paradigm \cite{Hadfield2019AOA,Fuchs2022ConstrainedMixers,
tsvelikhovskiy2024symmetries}. 
A depth-\(p\) CE--QAOA circuit prepares
\begin{equation}
\label{eq:ce-qaoa}
\ket{\psi_p(\vec\gamma,\vec\beta)}
=
\left(
\prod_{\ell=1}^{p}
U_M(\beta_\ell)e^{-i\gamma_\ell H_C}
\right)
\ket{s_0},
\qquad
U_M(\beta)
=
\exp\!\left(
-\frac{i\beta}{2}
\sum_{b=1}^{m}
\widetilde H_{XY}^{(b)}
\right).
\end{equation}

\item \emph{Problem-dependent classical maps. Geometry-aware classical primitives}, which read the
constraint-violation pattern of a measured sample and deterministically map that sample into the feasible set. Because \(U_M(\beta)\) preserves the fixed-weight sector and \(H_C\) is
diagonal, every layer maps the encoded subspace to itself:
\[
\ket{\psi_p(\vec\gamma,\vec\beta)}\in\OH.
\]
Measurement therefore produces a Born law
\[
p(x)
=
\left|
\langle x|\psi_p(\vec\gamma,\vec\beta)\rangle
\right|^2,
\qquad
x\in\mathcal X.
\]
The mixer controls transport on \(\mathcal X\), while the diagonal phases
control interference among the transported amplitudes. Neither operation,
however, requires every measured string to lie in the smaller globally
feasible set \(\mathcal F\subseteq\mathcal X\).

\end{enumerate}

The first three primitives provide the necessary ingredients for generating the
structured quantum sampling law. The fifth primitive, the classical map,
continues to exploit problem-derived structural information retained in the
quantum-generated samples, extending problem--algorithm co-design beyond
quantum measurement.

\subsection{Application classes for problem-dependent classical maps}
\label{subsec:classical-map-applications}

The hard constraints appearing in constrained optimization problems
typically give rise to multidimensional combinatorial data%
~\cite{AchlioptasCojaOghlan2008,DiaconisSturmfels1998,
DeLoeraOnn2006,SlavkovicZhuPetrovic2015}.
Assignment constraints generate row and column populations, routing constraints
generate degree, visit, load, and connectivity statistics, scheduling
constraints generate machine-load and conflict profiles, and coloring
constraints generate local conflict patterns. These statistics enter the
quantum formulation through the residual terms used to construct the penalty
Hamiltonian. The scalar value of \(H_{\mathrm{pen}}\) aggregates this
multidimensional constraint information. Examining its constituent residuals
therefore exposes structural statistics associated with each measured sample which can be used by a problem-dependent classical map to restore
feasibility while retaining useful bias present in the sampled output.
We specify this structure next.

\subsubsection{Multidimensional combinatorial data}

Let
\[
T:\mathcal X\longrightarrow\mathcal T
\]
denote the structural statistic inherited from the constraint formulation
underlying the penalty Hamiltonian in
Def.~\ref{def:kernel-requirement}. Its coordinates may contain marginal
counts, degree residuals, collision multiplicities, capacity excesses, visit
populations, conflict locations, machine loads, or related combinatorial
information. Suppose that the penalty depends on this statistic through
\begin{equation}
\label{eq:scheme-alignment-present2}
H_{\mathrm{pen}}(x)=\Phi(T(x)).
\end{equation}
Then every penalty level
\[
L_s
=
\left\{
x\in\mathcal X
\,\middle|\,
H_{\mathrm{pen}}(x)=s
\right\}
\]
is a union of \(T\)-fibres,
\[
L_s
=
\bigcup_{\tau\in\mathcal T\,:\,\Phi(\tau)=s}
L_\tau^T,
\qquad
L_\tau^T
=
\left\{
x\in\mathcal X
\,\middle|\,
T(x)=\tau
\right\}.
\]
The statistic \(T\) therefore resolves the combinatorial classes compressed
into each scalar penalty level in
Def.~\ref{def:kernel-requirement}.

A common penalty construction makes this relation explicit. Suppose that
\[
T(x)
=
\bigl(T_1(x),\ldots,T_q(x)\bigr)
\]
collects the residual statistics associated with the individual constraints,
and let the scalar aggregation map be
\[
\Phi(\tau)
=
\sum_{a=1}^{q}
\lambda_a\,\phi_a(\tau_a),
\qquad
\tau=(\tau_1,\ldots,\tau_q)\in\mathcal T,
\]
where the functions \(\phi_a\) penalize deviations from the required values.
Then
\[
H_{\mathrm{pen}}(x)
=
\Phi(T(x))
=
\sum_{a=1}^{q}
\lambda_a\,
\phi_a\bigl(T_a(x)\bigr).
\]
The scalar energy combines the coordinates of \(T(x)\), while the vector
\(T(x)\) retains the information needed to identify the corresponding
violation pattern. Evaluating the individual residuals after measurement
therefore provides a natural input for a deterministic repair operation.

For the product-space encodings studied here, a useful example is supplied by
transition-count statistics. Fix a reference configuration
\(y\in\mathcal X=[n]^m\) and define
\[
N^{(y)}_{u,v}(x)
=
\#\left\{
b\in\{1,\ldots,m\}
\,\middle|\,
y_b=u,\ x_b=v
\right\},
\qquad
u,v\in[n].
\]
The matrix \(N^{(y)}(x)\) may itself be regarded as a structural statistic,
\(
T_y(x)=N^{(y)}(x).
\)
Its column marginals recover the symbol populations of \(x\),
\[
h_v(x)
=
\sum_{u=1}^{n}
N^{(y)}_{u,v}(x).
\]
For a constraint requiring prescribed symbol multiplicities
\(c_1,\ldots,c_n\), define
\[
\Phi(N)
=
\sum_{v=1}^{n}
\lambda_v
\left(
\sum_{u=1}^{n}N_{u,v}-c_v
\right)^2.
\]
The corresponding penalty is then
\[
H_{\mathrm{pen}}(x)
=
\Phi\!\left(T_y(x)\right)
=
\Phi\!\left(N^{(y)}(x)\right)
=
\sum_{v=1}^{n}
\lambda_v
\bigl(h_v(x)-c_v\bigr)^2.
\]
For permutation constraints with \(m=n\), one has \(c_v=1\) for every
\(v\), giving
\[
H_{\mathrm{pen}}(x)
=
\sum_{v=1}^{n}
\lambda_v
\left(
\sum_{u=1}^{n}N^{(y)}_{u,v}(x)-1
\right)^2.
\]
Thus the scalar penalty is obtained by applying \(\Phi\) to the
multidimensional transition-count statistic, while
\(N^{(y)}(x)\) retains the finer information specifying how the constraint
violation is distributed among the symbols. The row sums of \(N^{(y)}(x)\) are fixed by the symbol populations of the
reference configuration,
\[
\sum_{v=1}^{n}
N^{(y)}_{u,v}(x)
=
\#\{b\mid y_b=u\},
\]
and the column sums recover the symbol histogram of \(x\),
\[
h_v(x)
=
\sum_{u=1}^{n}
N^{(y)}_{u,v}(x).
\]
The Hamming distance is obtained from the diagonal through
\[
d(x,y)
=
m-\operatorname{tr}N^{(y)}(x).
\]
The matrix \(N^{(y)}(x)\) retains the complete transition-count profile
between \(y\) and \(x\). Its diagonal determines the shell index, its column
marginals determine the symbol populations of \(x\), and its off-diagonal
entries identify the types and multiplicities of the local substitutions.
Many constraints involving symbol balance, assignment collisions, capacities,
visit multiplicities, and permutation conditions can therefore be expressed
through the marginal and contingency-table data contained in
\(N^{(y)}(x)\)%
~\cite{DiaconisSturmfels1998,SlavkovicZhuPetrovic2015}.

The same principle appears throughout classical constrained optimization.
Fixed-margin contingency-table fibres are defined through prescribed
marginals%
~\cite{DiaconisSturmfels1998,DeLoeraOnn2006,
SlavkovicZhuPetrovic2015}. Universality results for integer programming are
expressed through slim three-way transportation programs%
~\cite{DeLoeraOnn2006}. Overlap matrices and local conflict
profiles organize random coloring and constraint-satisfaction problems%
~\cite{AchlioptasCojaOghlan2008,CojaOghlanVilenchik2016}. Routing,
assignment, scheduling, allocation, and packing formulations likewise contain
multidimensional statistics that identify and characterize their constraint
violations%
~\cite{MnichWiese2015,KnopKoutecky2018,CosmadakisPapadimitriou1984}.
Further examples of these constraint structures are discussed in
App.~\ref{app:problem-stat}. The following subsection formalizes the
connection between constraint geometry, structural statistics, and
feasibility repair.

\subsection{Constraint geometry and feasibility repair}
As demonstrated in the foregoing, the scalar penalty value aggregates the multidimensional geometry of a
constraint violation. We therefore introduce a problem-dependent structural
statistic
\[
T:\mathcal X\longrightarrow\mathcal T
\]
that resolves the constituent violation pattern. %Where 
The statistic \(T\) refines the penalty levels in the sense that
\begin{equation}
\label{eq:penalty-factor-through-statistic}
H_{\mathrm{pen}}(x)
=
\Phi\bigl(T(x)\bigr)
\end{equation}
for some function \(\Phi\). Thus, two configurations with the same scalar
penalty can still be distinguished by the particular violation patterns
that produced that penalty.

Let
\[
\mathcal F
=
\{x\in\mathcal X:H_{\mathrm{pen}}(x)=0,\;
H'_{\mathrm{pen}}(x)=0\}
\]
denote the globally feasible set. A deterministic feasibility-repair
primitive is a map
\[
R:\mathcal X\longrightarrow\mathcal F
\]
satisfying
\begin{equation}
\label{eq:repair-retraction}
R(x)=x,
\qquad
x\in\mathcal F.
\end{equation}
In general, the correction can depend on the complete violation signature,
\[
R(x)=R_{T(x)}(x),
\]
so that \(T(x)\) identifies the relevant constraint geometry and
\(R_{T(x)}\) applies the corresponding correction. A particularly important realization is a metric projection. Given a
problem-compatible distance \(d_{\mathcal X}\), define
\begin{equation}
\label{eq:geometry-repair}
R_d(x)
\in
\operatorname*{arg\,min}_{y\in\mathcal F}
d_{\mathcal X}(x,y),
\end{equation}
with a fixed deterministic rule for resolving ties. The metric is defined directly on the encoded word space and weighted to reflect the combinatorial meaning of different violations.
%
% \sh{The existence of this projection does not imply efficient
% computability. Beyond permutation repair, efficient total repair
% requires problem-specific algorithms and feasibility assumptions.}

For permutation constraints with \(m=n\), the encoded space is
\(\mathcal X=[n]^n\) and the feasible set is identified with \(S_n\).
The Hamming-nearest repair is
\begin{equation}
\label{eq:hungarian-repair}
R_{\mathrm{Hun}}(x)
=
\min_{\prec}
\operatorname*{arg\,max}_{\pi\in S_n}
\sum_{i=1}^{n}
\mathbf 1\{x_i=\pi(i)\},
\end{equation}
where \(\prec\) is any fixed total ordering used to break ties. Equivalently, \(R_{\mathrm{Hun}}\) minimizes \(d_H(x,\pi)\) over feasible permutations. The corresponding maximum-overlap assignment is computed by the Hungarian algorithm in \(O(n^3)\) time \cite{Kuhn1955Hungarian}.

\section{Pushforward Law and Repair Basins}
\label{subsec:definition-repair-basins}

\subsection{Pushforward law and repair basins}
We now record the basic identities that govern deterministic repair as a pushforward of the raw quantum sampling law. Throughout this section, let \(\mathcal X\) be the raw sample space, let \(F\subseteq \mathcal X\) be the feasible set, and let \(R:\mathcal X\to F\) be a deterministic repair map. Let \(p\) be a probability law on \(\mathcal X\), and let \(p_R=R_\#p\) denote the repaired feasible law induced by \(R\). The additional structural properties of deterministic repair, including normalization, functoriality, idempotency, compositional closure, total-variation contraction, and efficient samplewise realization, are collected in Prop.~\ref{prop:pushforward-properties} of
App.~\ref{app:exact-pushforward-properties}.

\begin{proposition}[Pushforward law and repair basins]
\label{prop:pushforward-law}
For every feasible solution \(y\in F\), the repaired law satisfies
\[
p_R(y)=\sum_{x:\,R(x)=y} p(x).
\]
Equivalently, if
\[
B_y:=\{x\in \mathcal X:\,R(x)=y\}
\]
denotes the repair basin of \(y\), then
\[
p_R(y)=\sum_{x\in B_y} p(x).
\]
\end{proposition}

\begin{proof}
By definition, the repaired random variable is \(Y=R(Z)\), where \(Z\sim p\) on \(\mathcal X\). Therefore,
\[
p_R(y)
=
\Pr[Y=y]
=
\Pr[R(Z)=y].
\]
Since \(R\) is deterministic, the event \(\{R(Z)=y\}\) is exactly the disjoint union of the events \(\{Z=x\}\) over all \(x\in \mathcal X\) such that \(R(x)=y\). Hence
\[
\Pr[R(Z)=y]
=
\sum_{x:\,R(x)=y} \Pr[Z=x]
=
\sum_{x:\,R(x)=y} p(x),
\]
which is the desired identity. Rewriting the index set as the repair basin \(B_y\) gives the equivalent basin form
\[
p_R(y)=\sum_{x\in B_y} p(x).
\]
\end{proof}

\begin{proposition}[Repaired expected score]
\label{prop:repaired-expected-score}
Let \(C:F\to\mathbb R\) be any score or cost function on the feasible set. Then
\[
\mathbb E_p[C(R(Z))]
=
\sum_{x\in \mathcal X}(C\circ R)(x)\,p(x).
\]
Equivalently,
\[
\mathbb E_{p_R}[C]
=
\sum_{y\in F} C(y)\,p_R(y).
\]
\end{proposition}

\begin{proof}
Since \(Z\sim p\) on the discrete space \(\mathcal X\), the expectation of the random variable \(C(R(Z))\) is given by
\[
\mathbb E_p[C(R(Z))]
=
\sum_{x\in \mathcal X} C(R(x))\,p(x).
\]
By definition of composition, \(C(R(x))=(C\circ R)(x)\), so
\[
\mathbb E_p[C(R(Z))]
=
\sum_{x\in \mathcal X}(C\circ R)(x)\,p(x).
\]
This proves the first identity.

For the second form, note that \(Y=R(Z)\) has law \(p_R\) on \(F\), so
\[
\mathbb E_{p_R}[C]
=
\sum_{y\in F} C(y)\,p_R(y),
\]
which is the usual expectation formula under the repaired law.
\end{proof}

\subsection{Success probability after repair}
\label{subsec:success-probability-after-repair}

Let \(G\subseteq F\) be any set of good feasible solutions. Depending on the application, \(G\) may be the set of optimal solutions, the set of solutions below a target cost threshold, or the set of solutions achieving a desired approximation guarantee. The probability of success after repair is then
\[
\Pr(R(Z)\in G).
\]
Since \(R(Z)\) has law \(p_R\), this may be written as
\[
\Pr(R(Z)\in G)
=
\sum_{y\in G} p_R(y).
\]
Using again the pushforward structure, we may rewrite this in terms of the raw law \(p\):
\[
\Pr(R(Z)\in G)
=
\sum_{y\in G} p_R(y)
=
\sum_{x:\,R(x)\in G} p(x).
\]
Equivalently,
\[
\Pr(R(Z)\in G)
=
p\bigl(R^{-1}(G)\bigr),
\]
where
\[
R^{-1}(G)
=
\{x\in \mathcal X:\,R(x)\in G\}
\]
is the full preimage of the good set under the repair map. The following result holds.

\begin{proposition}[Repaired success probability]
\label{prop:repaired-success-probability}
Let \(G\subseteq F\) be any target set of feasible solutions, for example the optimal set or an approximation class. Then
\[
\Pr_p(R(Z)\in G)=p(R^{-1}(G)).
\]
Equivalently,
\[
\Pr_p(R(Z)\in G)
=
\sum_{x\in R^{-1}(G)} p(x).
\]
\end{proposition}

\begin{proof}
By definition,
\[
\Pr_p(R(Z)\in G)
=
\Pr_p\bigl(\{z\in \mathcal X:\,R(z)\in G\}\bigr).
\]
But the event \(\{z\in \mathcal X:\,R(z)\in G\}\) is precisely the preimage \(R^{-1}(G)\). Therefore
\[
\Pr_p(R(Z)\in G)
=
p(R^{-1}(G)).
\]
Since \(\mathcal X\) is discrete, this may also be written as
\[
p(R^{-1}(G))
=
\sum_{x\in R^{-1}(G)} p(x),
\]
which gives the second form.
\end{proof}

Applying Proposition~\ref{prop:repaired-success-probability} to two raw sampling laws
\(p\) and \(u\) gives
\[
\Pr_{p}(R(Z)\in G)-\Pr_{u}(R(W)\in G)
=
\sum_{x\in R^{-1}(G)}
\bigl(p(x)-u(x)\bigr),
\]
so the difference in repaired success probabilities is the excess
probability mass that \(p\) assigns to the repair preimage \(R^{-1}(G)\)
relative to \(u\). That is, the repaired law aggregates the probability of
all raw outcomes that are mapped into the target set \(G\). This is the
mechanism by which deterministic repair can harvest latent instance-specific
structure from the raw quantum sampling distribution. The pushforward formulation itself is sampler-independent and applies equally
to full-space samplers such as generic QAOA and to structured subspace samplers
such as CE--QAOA. The corresponding full-space specialization for generic
QAOA is discussed in App.~\ref{subsec:generic-qaoa}.

% Let \(p\) and \(u\) be two probability laws on the same raw sample space \(\mathcal X\), and let \(G\subseteq F\) be a target set of feasible solutions. Then
% \[
% \Pr_p(R(Z)\in G)-\Pr_u(R(W)\in G)
% =
% \sum_{x\in R^{-1}(G)} (p(x)-u(x)),
% \]
% where \(Z\sim p\) and \(W\sim u\). We show this by applying Prop. \ref{prop:repaired-success-probability} separately to the laws \(p\) and \(u\), we obtain
% \[
% \Pr_p(R(Z)\in G)=p(R^{-1}(G))
% \qquad\text{and}\qquad
% \Pr_u(R(W)\in G)=u(R^{-1}(G)).
% \]
% Subtracting these identities yields
% \[
% \Pr_p(R(Z)\in G)-\Pr_u(R(W)\in G)
% =
% p(R^{-1}(G)) - u(R^{-1}(G)).
% \]
% Since \(\mathcal X\) is discrete,
% \[
% p(R^{-1}(G)) - u(R^{-1}(G))
% =
% \sum_{x\in R^{-1}(G)} p(x)
% -
% \sum_{x\in R^{-1}(G)} u(x)
% =
% \sum_{x\in R^{-1}(G)} (p(x)-u(x)),
% \]
% which proves the claim.

% Applying Proposition~\ref{prop:repaired-success} to two raw sampling laws
% \(p\) and \(u\) gives
% \[
% \Pr_{p}(R(Z)\in G)-\Pr_{u}(R(W)\in G)
% =
% \sum_{x\in R^{-1}(G)}
% \bigl(p(x)-u(x)\bigr),
% \]
% so the difference in repaired success probabilities is the excess
% probability mass that \(p\) assigns to the repair preimage \(R^{-1}(G)\)
% relative to \(u\). i.e. the repaired law reveals the full channel by summing over all raw outcomes that are mapped to that solution. This is the mechanism by which deterministic repair can harvest latent instance-specific structure from the raw quantum sampling distribution. The pushforward formulation itself is sampler-independent and applies equally to full-space samplers such as generic QAOA and to structured subspace samplers such as CE--QAOA. 

\subsection{Noise, leakage, and breakdown of encoded basin compensation}
\label{subsec:noise-leakage-basin-breakdown}
Recall the CE-QAOA sampler defined by the circuit in Eq. \ref{eq:ce-qaoa}. In the noiseless CE--QAOA setting, the raw law is supported on the encoded subspace
\[
\mathcal X=[n]^n\subset \Omega:=\{0,1\}^{n^2},
\]
rather than on the full ambient hypercube \(\Omega\). 
For permutation-constrained problems, the repair basins
partition the encoded space $\mathcal X=[n]^n$, giving
\[
\frac{1}{n!}\sum_{\pi\in S_n}
|R^{-1}(\pi)\cap\mathcal X|
=\frac{n^n}{n!}.
\]
For a fixed optimal permutation $\pi^\star$, however,
the corresponding basin has cardinality
\[
|A_\star|
=|R^{-1}(\pi^\star)\cap\mathcal X|
=\kappa_\star\frac{n^n}{n!},
\]
where $\kappa_\star$ measures its size relative to
the average. The repaired optimum probability is
therefore governed by both the optimal-basin volume
and the quantum probability concentrated within it.
At the average-basin scale $\kappa_\star=1$, the
basin volume compensates the encoded normalization
$n^{-n}$, producing the factorial scale $1/n!$.\footnote{Throughout, we use \(f(n)\asymp g(n)\) to denote equality up to positive
multiplicative constants, equivalently \(f(n)=\Theta(g(n))\). We write
\(f(n)\gtrsim g(n)\) when \(f(n)\) is bounded below by \(g(n)\) up to a
positive multiplicative constant. These asymptotic relations are understood
in the relevant large-instance limit unless stated otherwise.
}

% The repaired optimum probability is governed by transport inside the encoded subspace, so the factor \(n^n\) is compensated by the basin size and replaced by the smaller feasible-subspace scale \(n!\).

On noisy hardware, however, this mechanism can break down. In particular, coherent or incoherent errors may leak probability mass outside the encoded subspace and into the ambient cube \(\Omega\setminus \mathcal X\). Once this happens, the average basin factor \(|A_\star|\asymp n^n/n!\) no longer controls the full repaired law, because part of the repaired output is now inherited from raw samples that never belonged to the encoded subspace in the first place. We formalize this by separating the noisy raw law into its encoded and leaked components.

\begin{definition}[Leakage mass and sector decomposition]
\label{def:leakage-mass}
Let \(p_\lambda\) be a noisy raw sampling law on the ambient outcome space \(\Omega\), where \(\lambda\) denotes the noise strength. Define the leakage mass outside the encoded subspace by
\[
\eta_\lambda
:=
p_\lambda(\Omega\setminus \mathcal X).
\]
Whenever \(0<\eta_\lambda<1\), define the conditional laws
\[
p_\lambda^{\mathcal X}(x)
:=
\frac{p_\lambda(x)}{1-\eta_\lambda},
\qquad x\in \mathcal X,
\]
and
\[
p_\lambda^{\mathrm{out}}(x)
:=
\frac{p_\lambda(x)}{\eta_\lambda},
\qquad x\in \Omega\setminus \mathcal X.
\]
Then
\begin{equation}
\label{eq:sector-decomposition}
p_\lambda
=
(1-\eta_\lambda)\,p_\lambda^{\mathcal X}
+
\eta_\lambda\,p_\lambda^{\mathrm{out}},
\end{equation}
where \(p_\lambda^\mathcal X\) is supported on \(\mathcal X\) and \(p_\lambda^{\mathrm{out}}\) is supported on \(\Omega\setminus \mathcal X\).
\end{definition}

The repaired law inherits the same convex decomposition.

\begin{proposition}[Pushforward decomposition under leakage]
\label{prop:pushforward-decomposition-leakage}
Let \(R:\Omega\to F\) be a deterministic repair map into the feasible set \(F\subseteq \mathcal X\). Then the repaired noisy law satisfies
\begin{equation}
\label{eq:pushforward-decomposition-leakage}
R_\# p_\lambda
=
(1-\eta_\lambda)\,R_\# p_\lambda^{\mathcal X}
+
\eta_\lambda\,R_\# p_\lambda^{\mathrm{out}}.
\end{equation}
Equivalently, for every feasible \(y\in F\),
\[
(R_\# p_\lambda)(y)
=
(1-\eta_\lambda)(R_\# p_\lambda^{\mathcal X})(y)
+
\eta_\lambda (R_\# p_\lambda^{\mathrm{out}})(y).
\]
\end{proposition}

\begin{proof}
Apply the linearity of pushforward to the decomposition
\eqref{eq:sector-decomposition}:
\[
R_\# p_\lambda
=
R_\#\!\left((1-\eta_\lambda)p_\lambda^{\mathcal X}
+
\eta_\lambda p_\lambda^{\mathrm{out}}\right)
=
(1-\eta_\lambda)R_\# p_\lambda^{\mathcal X}
+
\eta_\lambda R_\# p_\lambda^{\mathrm{out}}.
\]
The pointwise formula is the corresponding statement evaluated on singleton feasible outputs.
\end{proof}

Thus repaired success against any target set \(G\subseteq F\) decomposes into an encoded contribution and a leaked contribution. That is, let \(G\subseteq F\) be any target set of feasible solutions. Then
\begin{equation}
\label{eq:repaired-success-decomposition}
P^{\mathrm{rep}}_\lambda(G)
:=
\Pr_{p_\lambda}(R(Z)\in G)
=
(1-\eta_\lambda)\,P^{\mathrm{rep},\mathcal X}_\lambda(G)
+
\eta_\lambda\,P^{\mathrm{rep,out}}_\lambda(G),
\end{equation}
where
\[
P^{\mathrm{rep},\mathcal X}_\lambda(G)
:=
\Pr_{p_\lambda^{\mathcal X}}(R(Z)\in G),
\qquad
P^{\mathrm{rep,out}}_\lambda(G)
:=
\Pr_{p_\lambda^{\mathrm{out}}}(R(Z)\in G).
\]

The decomposition \eqref{eq:repaired-success-decomposition} shows precisely that only the fraction \(1-\eta_\lambda\) of the raw law that remains inside the encoded subspace can still benefit from the encoded-subspace basin geometry. To make this explicit, define the encoded good-preimage size and the leaked good-preimage size by
\[
|R^{-1}(G)\cap \mathcal X|
\qquad\text{and}\qquad
|R^{-1}(G)\setminus \mathcal X|.
\]
Equivalently, introducing the average basin scales
\[
\overline B_G^{\mathcal X}
:=
\frac{|R^{-1}(G)\cap \mathcal X|}{|G|},
\qquad
\overline B_G^{\mathrm{out}}
:=
\frac{|R^{-1}(G)\setminus \mathcal X|}{|G|},
\]
we may write the repaired success in the form
\begin{equation}
\label{eq:noisy-basin-factorization}
P^{\mathrm{rep}}_\lambda(G)
=
(1-\eta_\lambda)\,
c_G^{\mathcal X}(\lambda)\,
\overline B_G^{\mathcal X}\,
\frac{|G|}{|\mathcal X|}
+
\eta_\lambda\,
c_G^{\mathrm{out}}(\lambda)\,
\overline B_G^{\mathrm{out}}\,
\frac{|G|}{|\Omega\setminus \mathcal X|},
\end{equation}
where the concentration coefficients are defined by
\[
c_G^{\mathcal X}(\lambda)
:=
\frac{p_\lambda^{\mathcal X}(R^{-1}(G)\cap \mathcal X)}{|R^{-1}(G)\cap \mathcal X|/|\mathcal X|},
\qquad
c_G^{\mathrm{out}}(\lambda)
:=
\frac{p_\lambda^{\mathrm{out}}(R^{-1}(G)\setminus \mathcal X)}{|R^{-1}(G)\setminus \mathcal X|/|\Omega\setminus \mathcal X|}.
\]

For the singleton optimum target \(G=\{\pi^\star\}\), this becomes
\begin{equation}
\label{eq:noisy-optimum-basin-factorization}
P^{\mathrm{rep}}_\lambda(\pi^\star)
=
(1-\eta_\lambda)\,
c_\star^{\mathcal X}(\lambda)\,
\frac{|R^{-1}(\pi^\star)\cap \mathcal X|}{|\mathcal X|}
+
\eta_\lambda\,
c_\star^{\mathrm{out}}(\lambda)\,
\frac{|R^{-1}(\pi^\star)\setminus \mathcal X|}{|\Omega\setminus \mathcal X|}.
\end{equation}

If $R^{-1}(G)\setminus X=\varnothing$, the out-of-sector contribution in Eq.~\ref{eq:noisy-basin-factorization} is defined to be zero and
$c_G^{\mathrm{out}}(\lambda)$ need not be introduced. Otherwise,
$c_G^{\mathrm{out}}(\lambda)$ is given by the preceding ratio. 
Using the optimal-basin cardinality
\(
|R^{-1}(\pi^\star)\cap\mathcal X|
=\kappa_\star\frac{n^n}{n!},
\qquad |\mathcal X|=n^n,
\)
the encoded contribution in Eq.~(13) is exactly
\(
(1-\eta_\lambda)c_\star^{\mathcal X}(\lambda)
\frac{\kappa_\star}{n!}.
\)
At the average-basin scale $\kappa_\star=1$,
this reduces to the factorial normalization
$1/n!$, multiplied by the surviving encoded-sector
weight and its concentration factor.

Thus the noiseless compensation of the ambient encoded factor \(n^n\) survives only on the non-leaked weight \(1-\eta_\lambda\). As leakage grows, the encoded-subspace basin-transfer mechanism is linearly suppressed. Equation~\eqref{eq:noisy-optimum-basin-factorization} isolates the two distinct ways in which noise can destroy pushforward advantage, namely, through sector-violating leakage and in-sector concentration loss. Even if leakage is small, noise can reduce the encoded concentration factor \(c_\star^{\mathcal X}(\lambda)\) by washing out the constructive interference or shell-aligned mass concentration that originally favored the good repair preimage. The first mechanism is geometric and the second mechanism is dynamical. Coherent noise can be especially harmful because it usually generates both at once.

\section{Case study: TSP with deterministic repair}
\label{sec:case-study-tsp}

\subsection{TSP specialization of the  OFM kernel}
\label{subsec:tsp-encoding}
This case study is natural for two reasons. First, the TSP admits a double one-hot permutation encoding compatible with the  OFM kernel. Second, deterministic repair from the encoded row-one-hot subspace to
the feasible permutation sector is obtained by solving a
maximum-overlap assignment problem with the Hungarian
algorithm~\cite{Kuhn1955Hungarian}.

Let \(G=(V,E)\) be a complete weighted graph with \(|V|=n\) and distance matrix
\(C\in\mathbb R_{\ge 0}^{n\times n}\). A tour is represented by an \(n\times n\)
binary assignment matrix \( P=(x_{i,a})\), where \(i\) denotes the tour position
and \(a\) the visited city. Feasible tours satisfy the double one-hot
constraints
\begin{equation}
\sum_{a=1}^{n}x_{i,a}=1\quad(\forall i),
\qquad
\sum_{i=1}^{n}x_{i,a}=1\quad(\forall a),
\qquad
x_{i,a}\in\{0,1\}.
\end{equation}
This matches the  OFM kernel with \(m=n\) one-hot blocks, one per tour
position, each of local size \(n\).

On the computational basis \(x\mapsto \mathcal P =(x_{i,a})\), the objective is
\begin{equation}
\label{eq:tsp-obj-case-study-clean}
H_{\mathrm{obj}}(x)
=
\sum_{i=1}^{n}
\sum_{a,b=1}^{n}
C_{ab}\,x_{i,a}x_{i+1,b},
\qquad
x_{n+1,b}:=x_{1,b}.
\end{equation}
and feasibility is enforced by row- and column-sum penalties,
\begin{equation}
\label{eq:tsp-pen-case-study-clean}
H_{\mathrm{pen}}(x)
=
\lambda_{\mathrm{row}}\sum_{i=1}^{n}\Bigl(\sum_{a=1}^{n}x_{i,a}-1\Bigr)^{2}
+
\lambda_{\mathrm{col}}\sum_{a=1}^{n}\Bigl(\sum_{i=1}^{n}x_{i,a}-1\Bigr)^{2},
\end{equation}
optionally together with linear forbids. In the  OFM kernel, the row
one-hot conditions are already enforced dynamically by the encoded subspace and
the block-XY mixer, so on the actual sampled space the row penalty is
redundant. The remaining column constraints distinguish feasible tours from
infeasible encoded outputs.

Accordingly, the raw CE--QAOA outcome space is the encoded row-one-hot subspace
\begin{equation}
\label{eq:tsp-raw-space-clean}
\mathcal X
:=
\Bigl\{
M\in\{0,1\}^{n\times n}:\;
\sum_{a=1}^{n}M_{i,a}=1\ \text{for all }i
\Bigr\}
\cong [n]^n,
\end{equation}
while the feasible set is the permutation sector
\begin{equation}
\label{eq:tsp-feasible-set-clean}
F= S_n
:=
\Bigl\{
P\in \mathcal X:\;
\sum_{i=1}^{n}P_{i,a}=1\ \text{for all }a
\Bigr\}.
\end{equation}
Thus, infeasibility within $X$ arises through column collisions,
omissions, or both. Deterministic repair acts by mapping outputs to a feasible permutation matrix of maximum overlap, thereby inducing the repaired law \(
p_R=R_\# p\) on \(S_n\). Maximum-overlap repair can admit multiple optimal
assignments. A fixed deterministic tie-breaking rule
defines the repair map $R$, and the numerical
comparisons use the same implementation for every
sampler. Appendix~\ref{app:fair-repair} also constructs
a symmetry-controlled stochastic repair whose
pushforward of the encoded-uniform law is uniform
over $S_n$.

\subsection{Enhanced Sampling Advantage with Pushforwards}
\label{sec:subspace}

The shell-transfer formalism of
Ref.~\cite{onahseperating} expresses the amplitude
at a target configuration as a sum of contributions
from paths through the encoded subspace.
The corresponding absolute path weight is determined
by the mixer, while the phase-aligned bound quantifies
its coherent contribution to the target amplitude. We extend this analysis to the complete repair basin
of an optimal solution. 
% For the TSP, 
% let
% \(\mathcal X=[n]^n\), let \(S_n\) be the feasible
% permutation sector, and let
% \[
% R:\mathcal X\longrightarrow S_n
% \]
% be the deterministic Hungarian maximum-overlap
% repair map. Fix an optimal permutation
% \(\pi^\star\in S_n\) and define its repair basin
% \[
% A_\star:=R^{-1}(\pi^\star).
% \]

% The accumulated absolute path weight is \eqref{eq:v0-closed-form}
% where
% \[
% q_n(\beta)
% =
% |a_0(\beta)|+(n-1)|a_1(\beta)|,
% \]
% and, from Ref.~\cite[Lem.~2]{onahseperating},
% \[
% \begin{aligned}
% a_0(\beta)
% &=
% \frac{1}{n}e^{-i\beta}
% +
% \left(1-\frac{1}{n}\right)e^{i\beta/(n-1)},\\
% a_1(\beta)
% &=
% \frac{1}{n}
% \left(e^{-i\beta}-e^{i\beta/(n-1)}\right).
% \end{aligned}
% \]

% The repaired-success analysis follows by extending
% the pointwise phase-aligned estimate underlying
% Theorem~10 of Ref.~\cite{onahseperating} to every
% configuration in \(A_\star\).

%========================================================
\paragraph{Contributor viewpoint on repaired success}
\label{subsec:shell-transfer-pushforward-application}
%========================================================

In the shell-transfer formalism developed in
Ref.~\cite{onahseperating}, for a fixed target output
\(y\), the amplitude
\(\langle y\mid\psi_p(\vec\gamma,\vec\beta)\rangle\)
is built from many multi-step paths through the
encoded subspace. The shell-transfer recursion
packages the absolute weighted contribution of
those paths according to their distance from \(y\),
while the phase-alignment hypothesis determines
how much of that absolute mass survives in the
complex sum. Consequently, the shell-transfer quantities provide a count of potential contributors, and the phase theorem quantifies the survival of quantum amplitude.

To specialize to the TSP setting,
let \(m=n\), so \(\mathcal X=[n]^n\), the feasible
set is \(S_n\), and
\[
R:\mathcal X\to S_n
\]
is the deterministic Hungarian maximum-overlap
repair map. Fix an optimal permutation
\(\pi^\star\in S_n\), and let
\[
A_\star:=R^{-1}(\pi^\star)
\]
be its repair basin. The accumulated absolute amplitude weight is
\begin{equation}
\label{eq:v0-closed-form}
v_0^{(p)}
=
\prod_{\ell=1}^{p}q_n(\beta_\ell)^n,
\end{equation}
where
\[
q_n(\beta)
:=
|a_0(\beta)|+(n-1)|a_1(\beta)|,
\]
and, from Ref.~\cite[Lem.~2]{onahseperating},
\[
\begin{aligned}
a_0(\beta)
&=
\frac{1}{n}e^{-i\beta}
+
\left(1-\frac{1}{n}\right)
e^{i\beta/(n-1)},\\
a_1(\beta)
&=
\frac{1}{n}
\left(
e^{-i\beta}-e^{i\beta/(n-1)}
\right).
\end{aligned}
\]

The repair basins partition the encoded space,
giving
\[
\sum_{\pi\in S_n}|R^{-1}(\pi)|=n^n.
\]
Thus, the average repair-basin cardinality is
exactly \(n^n/n!\). We define the relative volume
of the optimal repair basin by
\begin{equation}
\label{eq:relative-optimal-basin-volume}
\kappa_\star
:=
\frac{n!|A_\star|}{n^n}.
\end{equation}
The average relative basin volume over \(S_n\)
is exactly one.

For the repaired-success analysis, one obtains
an adaptation of Theorem~10 in
Ref.~\cite{onahseperating} by extending the
pointwise phase-aligned estimate to the complete
repair basin.

\begin{corollary}[Pushforward enhanced overlap bound]
\label{cor:uniform-phase-aligned-repair}
Assume that for every \(y\in A_\star\), all
depth-\(p\) path summands ending at \(y\) lie in
a common arc of angular width \(\Theta_p<\pi\).
Then
\begin{equation}
\label{eq:repair-from-shell-transfer}
\begin{aligned}
P^{\mathrm{rep}}_{\mathrm{CE},p}(\pi^\star)
&:=
\sum_{y\in A_\star}
\bigl|\langle y\mid
\psi_p(\vec\gamma,\vec\beta)\rangle\bigr|^2\\
&\geq
\frac{|A_\star|}{n^n}
\cos^2\!\left(\frac{\Theta_p}{2}\right)
\bigl(v_0^{(p)}\bigr)^2.
\end{aligned}
\end{equation}
Using \eqref{eq:v0-closed-form}, this becomes
\begin{equation}
\label{eq:repair-from-shell-transfer-closed}
P^{\mathrm{rep}}_{\mathrm{CE},p}(\pi^\star)
\geq
\frac{|A_\star|}{n^n}
\cos^2\!\left(\frac{\Theta_p}{2}\right)
\prod_{\ell=1}^{p}q_n(\beta_\ell)^{2n}.
\end{equation}
Equivalently, in terms of the relative
optimal-basin volume,
\begin{equation}
\label{eq:repair-from-shell-transfer-average-basin}
P^{\mathrm{rep}}_{\mathrm{CE},p}(\pi^\star)
\geq
\frac{\kappa_\star}{n!}
\cos^2\!\left(\frac{\Theta_p}{2}\right)
\prod_{\ell=1}^{p}q_n(\beta_\ell)^{2n}.
\end{equation}
\end{corollary}

\begin{proof}
For each \(y\in A_\star\), the pointwise
phase-aligned bound of
Ref.~\cite[Theorem~8]{onahseperating},
specialized to \(m=n\), gives
\[
\bigl|\langle y\mid
\psi_p(\vec\gamma,\vec\beta)\rangle\bigr|^2
\geq
\frac{1}{n^n}
\cos^2\!\left(\frac{\Theta_p}{2}\right)
\bigl(v_0^{(p)}\bigr)^2.
\]
Summing this bound over the repair basin
\(A_\star\) yields
\[
\begin{aligned}
P^{\mathrm{rep}}_{\mathrm{CE},p}(\pi^\star)
&=
\sum_{y\in A_\star}
\bigl|\langle y\mid
\psi_p(\vec\gamma,\vec\beta)\rangle\bigr|^2\\
&\geq
\frac{|A_\star|}{n^n}
\cos^2\!\left(\frac{\Theta_p}{2}\right)
\bigl(v_0^{(p)}\bigr)^2.
\end{aligned}
\]
Using \eqref{eq:v0-closed-form} gives
\eqref{eq:repair-from-shell-transfer-closed}.
Finally,
\[
\frac{|A_\star|}{n^n}
=
\frac{\kappa_\star}{n!}
\]
establishes
\eqref{eq:repair-from-shell-transfer-average-basin}.
\end{proof}

The basin factor in Corollary~
\ref{cor:uniform-phase-aligned-repair}
aggregates the quantum probability across
all configurations repairing to \(\pi^\star\).
At the average basin scale
\(\kappa_\star=1\), the combinatorial volume
\(n^n/n!\) compensates the pointwise
normalization \(n^{-n}\), producing the
factorial scale \(1/n!\).
The relative basin volume \(\kappa_\star\)
quantifies the corresponding contribution
of the optimal repair basin. We now derive the resulting compensation
threshold and its finite-shot consequence.

\begin{theorem}[Enhanced depth threshold
from pushforward compensation]
\label{thm:log-depth-threshold}
Consider the OFM kernel TSP specialization
with \(n\geq4\) and encoded subspace
\(\mathcal X=[n]^n\).
Let
\[
A_\star:=R^{-1}(\pi^\star)
\]
be the repair basin of a fixed optimal
permutation, with relative basin volume
\[
\kappa_\star
=
\frac{n!|A_\star|}{n^n}.
\]

For any depth-\(p\) mixer-angle schedule
satisfying the phase-alignment hypothesis
of Corollary~
\ref{cor:uniform-phase-aligned-repair},
suppose that
\begin{equation}
\label{eq:pushforward-polynomial-shot-conditions}
\prod_{\ell=1}^{p}
q_n(\beta_\ell)^{2n}
\geq
\frac{n!}{\kappa_\star L^2},
\qquad
\cos\!\left(\frac{\Theta_p}{2}\right)
\geq\frac1K,
\end{equation}
where \(L=L(n)\geq1\) and \(K=K(n)\geq1\)
are polynomially bounded functions.
Then
\begin{equation}
\label{eq:pushforward-polynomial-success}
P^{\mathrm{rep}}_{\mathrm{CE},p}(\pi^\star)
\geq
\frac{1}{L^2K^2}.
\end{equation}
Consequently,
\begin{equation}
\label{eq:pushforward-shot-bound}
N=
\left\lceil
L^2K^2\log(1/\epsilon)
\right\rceil
\end{equation}
independent shots suffice to recover
\(\pi^\star\) with probability at least
\(1-\epsilon\).

For \(n\geq4\), define the maximal absolute
transfer factor
\begin{equation}
\label{eq:pushforward-qmax}
q_{\max}(n)
:=
\max_\beta q_n(\beta)
=
3-\frac4n,
\end{equation}
attained at
\(
\beta_\star
=
\frac{\pi(n-1)}n.
\)
The minimum depth at which the absolute
transfer factor can compensate the
optimal-basin normalization,
\begin{equation}
\label{eq:pushforward-compensation-condition}
\prod_{\ell=1}^{p}
q_n(\beta_\ell)^{2n}
\geq
\frac{n!}{\kappa_\star},
\end{equation}
is
\begin{equation}
\label{eq:pushforward-depth-threshold}
p_{\mathrm{abs}}(n,\kappa_\star)
=
\left\lceil
\frac{\log(n!/\kappa_\star)}
{2n\log q_{\max}(n)}
\right\rceil.
\end{equation}
\end{theorem}

\begin{proof}
By Corollary~
\ref{cor:uniform-phase-aligned-repair},
\[
P^{\mathrm{rep}}_{\mathrm{CE},p}(\pi^\star)
\geq
\frac{\kappa_\star}{n!}
\cos^2\!\left(\frac{\Theta_p}{2}\right)
\prod_{\ell=1}^{p}
q_n(\beta_\ell)^{2n}.
\]
The stated transport and phase conditions
give
\(
P^{\mathrm{rep}}_{\mathrm{CE},p}(\pi^\star)
\geq
\frac{1}{L^2K^2}.
\)
For \(N\) independent shots, the probability
of failing to recover \(\pi^\star\) satisfies
\[
\begin{aligned}
\Pr(\text{failure})
&=
\left(
1-P^{\mathrm{rep}}_{\mathrm{CE},p}(\pi^\star)
\right)^N \leq
e^{-N/(L^2K^2)}.
\end{aligned}
\]
The choice in
\eqref{eq:pushforward-shot-bound}
therefore establishes the finite-shot guarantee. By Ref.~\cite[Proposition~9]{onahseperating},
the maximum of \(q_n(\beta)\) is attained at
\[
\beta_\star=\frac{\pi(n-1)}n,
\qquad
q_n(\beta_\star)
=
q_{\max}(n)
=
3-\frac4n.
\]
For any mixer-angle schedule,
\[
\prod_{\ell=1}^{p}
q_n(\beta_\ell)^{2n}
\leq
q_{\max}(n)^{2np},
\]
with equality at
\(\beta_1=\cdots=\beta_p=\beta_\star\).
The compensation condition
\eqref{eq:pushforward-compensation-condition}
is reached at the minimum
depth satisfying
\(
q_{\max}(n)^{2np}
\geq
\frac{n!}{\kappa_\star}.
\)
Taking logarithms gives
\eqref{eq:pushforward-depth-threshold}.
\end{proof}

Note, however, that at the transport-maximizing schedule
%$$ \beta_\ell = \frac{\pi(n-1)}{n},
 %   \qquad \ell=1,\ldots,p, $$
 $ \beta_\ell = \frac{\pi(n-1)}{n}$,
 $\ell=1,\dots,p$,
Corollary~11 of Ref.~\cite{onahseperating} excludes the common-cone phase condition for $n\geq 4$ and $p\geq 2$. But this does not exclude useful sampling at other mixing angles or outside the common-cone regime. The depth $p_{\mathrm{abs}}$ characterizes
compensation of the absolute-transfer normalization
and, as already stated, the corresponding
repaired-success guarantee additionally requires
the phase-alignment hypothesis of
Corollary~\ref{cor:uniform-phase-aligned-repair}.
Indeed, since the repaired success probability
cannot exceed unity, the lower bound in
Corollary~\ref{cor:uniform-phase-aligned-repair}
implies the consistency condition
\[
\cos^2\left(\frac{\Theta_p}{2}\right)
\leq
\frac{n!}{
\kappa_\star
\displaystyle\prod_{\ell=1}^{p}
q_n(\beta_\ell)^{2n}
}.
\]
This condition quantifies the compatibility between
absolute transfer, phase alignment, and probability
normalization, consistent with the phase-interference
analysis of Ref.~\cite{onahseperating}.

% The absolute-transfer compensation threshold
% depends on the relative volume of the
% optimal repair basin:
% \[
% p_{\mathrm{abs}}(n,\kappa_\star)
% =
% \left\lceil
% \frac{\log n!-\log\kappa_\star}
% {2n\log(3-4/n)}
% \right\rceil.
% \]

% Thus, Hungarian pushforward repair contributes the combinatorial volume
% of the optimal repair basin to the absolute-transfer compensation factor.
% The correction \(-\log\kappa_\star\) quantifies its contribution to the reduced
% logarithmic-depth threshold. The phase-alignment requirements of the
% underlying shell-transfer bounds,
% including the transport-maximizing
% schedule, are analyzed in
% Ref.~\cite[Theorem~10 and
% Corollary~11]{onahseperating}.

\section{Numerical Study}
\label{sec:numerical-analysis}

We compare quantum and classical sampling laws on the same encoded
configuration space. For the TSP, this is the row-one-hot space
\(\mathcal X=[n]^n\), whose elements are represented by binary assignment
matrices with one active entry in each row. The classical reference sampler
draws one city label \(x_i\) independently and uniformly from \([n]\) for
each tour position \(i=1,\ldots,n\), and constructs
\[
    M_{i,a}=\mathbf 1\{x_i=a\},
    \qquad
    \sum_{a=1}^{n}M_{i,a}=1
    \quad\text{for every }i.
\]
Consequently, the classical reference law is
\begin{equation}
    u(x)=\frac{1}{n^n},\qquad x\in\mathcal X.
\end{equation}
It coincides exactly with the computational-basis measurement law of the
uniform CE--QAOA initial state \(|s_0\rangle\). The ideal CE--QAOA circuit
starts from this state and generates a quantum sampling law \(p\) on the
same row-one-hot space. The column constraints defining feasible
permutations are left to the problem Hamiltonian and the repair map.

We apply the same deterministic Hungarian maximum-overlap repair
\(R:\mathcal X\to S_n\) to both samplers and compare the resulting feasible
laws \(R_\#p\) and \(R_\#u\). Each instance uses equal numbers of quantum
and encoded-uniform samples. In particular, for any permutation
\(\pi\in S_n\), the classical reference probability is
\begin{equation}
    (R_\#u)(\pi)
    =\frac{|R^{-1}(\pi)|}{n^n}
    =\frac{\kappa_\pi}{n!},
    \qquad
    \kappa_\pi:=\frac{n!|R^{-1}(\pi)|}{n^n}.
\end{equation}
We define experimental success through the complete optimal-cost set
$$    C_\star := \min_{\pi\in S_n} C(\pi),
    \qquad
    G_\star := \{\pi\in S_n : C(\pi)=C_\star\}.$$
Thus, a raw sample $x$ is successful precisely when
$C(R(x))=C_\star$. The same event is used for every sampler. 

% For the encoded-uniform law $u(x)=n^{-n}$ and the implemented
% deterministic repair map $R$, the reference success probability is
% $$  (R_\#u)(G_\star) = \frac{|R^{-1}(G_\star)|}{n^n}.$$
% This probability depends on the implemented tie-breaking rule.
% We therefore use the actual good-set preimage probability rather
% than substituting an average single-permutation basin probability.}

\subsection{Noiseless Simulations}
\label{subsec:noiseless-analysis}

We study single-layer CE--QAOA on TSP instances with \(n=7,8,9,10\)
from Ref.~\cite{Osaba2024Qoptlib}, using the selected best angle for
each instance. Although the shell-transfer threshold results are
developed for exact lattice-normalized spectra in
Ref.~\cite{onahseperating}, earlier CE--QAOA analyses
\cite{onahfund,onahfinite} extend the phase-averaging treatment beyond exact lattice spectra through Riemann--Lebesgue averaging. Our numerics, obtained without lattice normalization, are consistent
with this broader interpretation. We now examine the resulting repaired sampling laws on some known benchmark instances.

\begin{figure*}[t]
    \centering
    \begin{subfigure}[t]{0.48\textwidth}
        \centering
        \includegraphics[width=\textwidth]{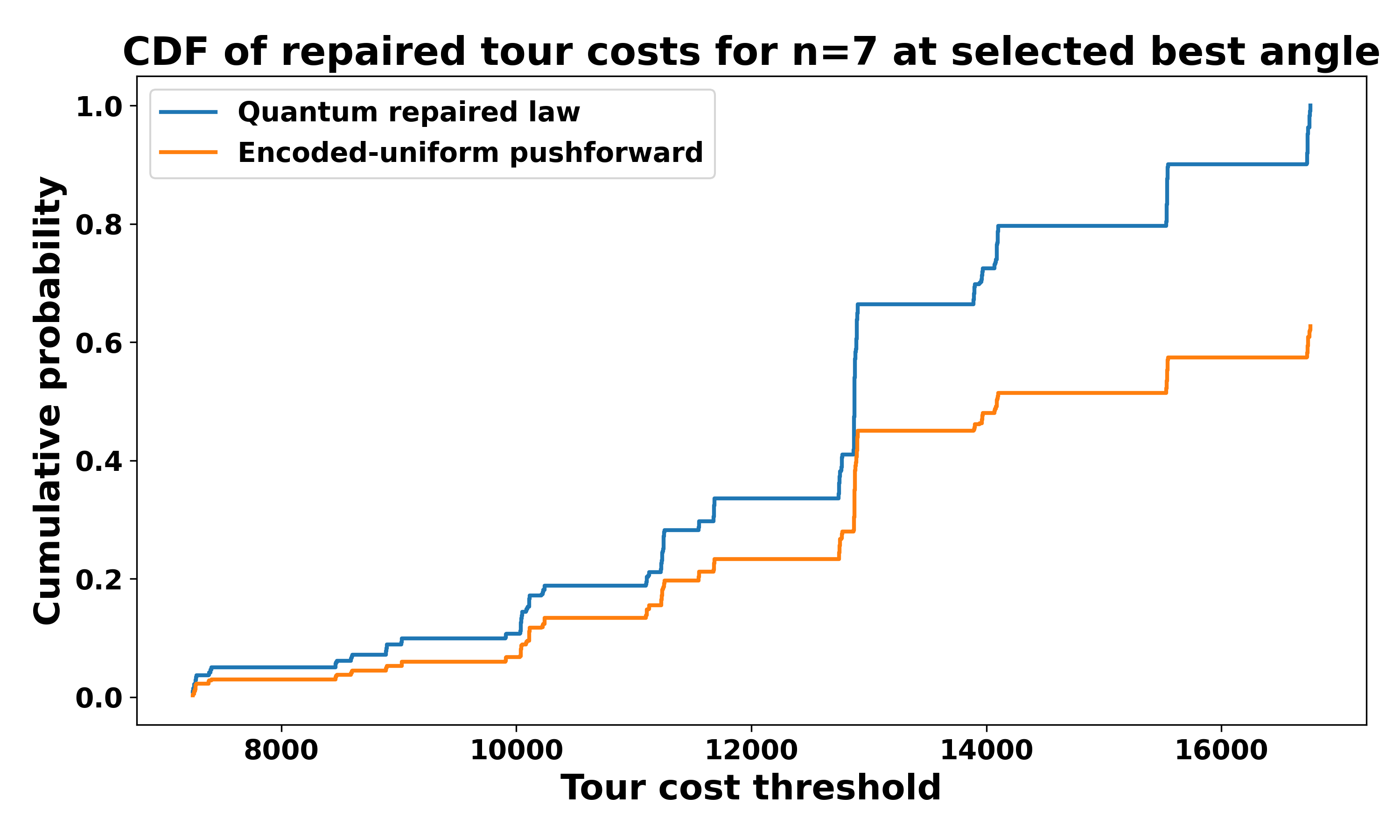}
        \caption{\(n=7\)}
        \label{fig:cdf-n7}
    \end{subfigure}
    \hfill
    \begin{subfigure}[t]{0.48\textwidth}
        \centering
        \includegraphics[width=\textwidth]{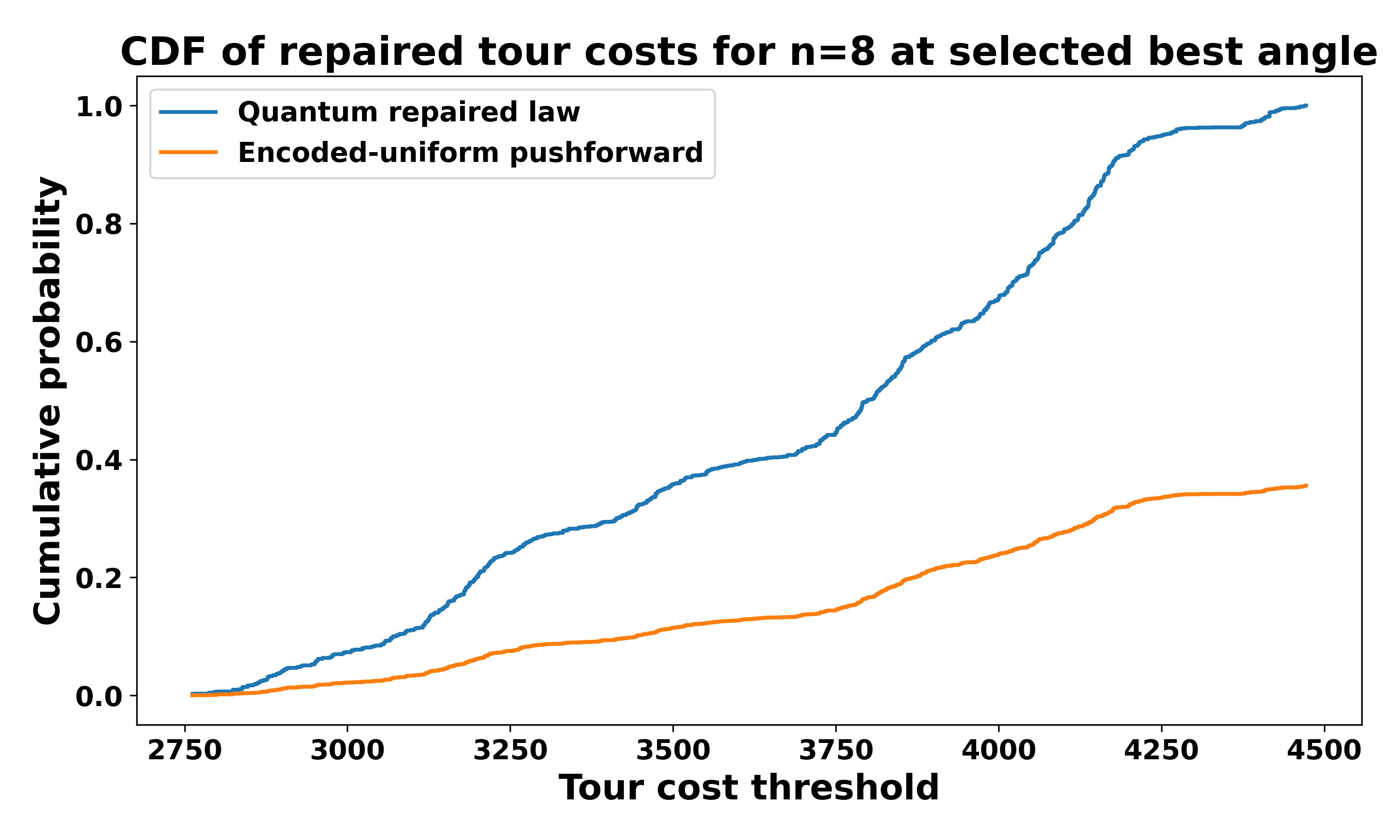}
        \caption{\(n=8\)}
        \label{fig:cdf-n8}
    \end{subfigure}

    \vspace{0.8em}

    \begin{subfigure}[t]{0.48\textwidth}
        \centering
        \includegraphics[width=\textwidth]{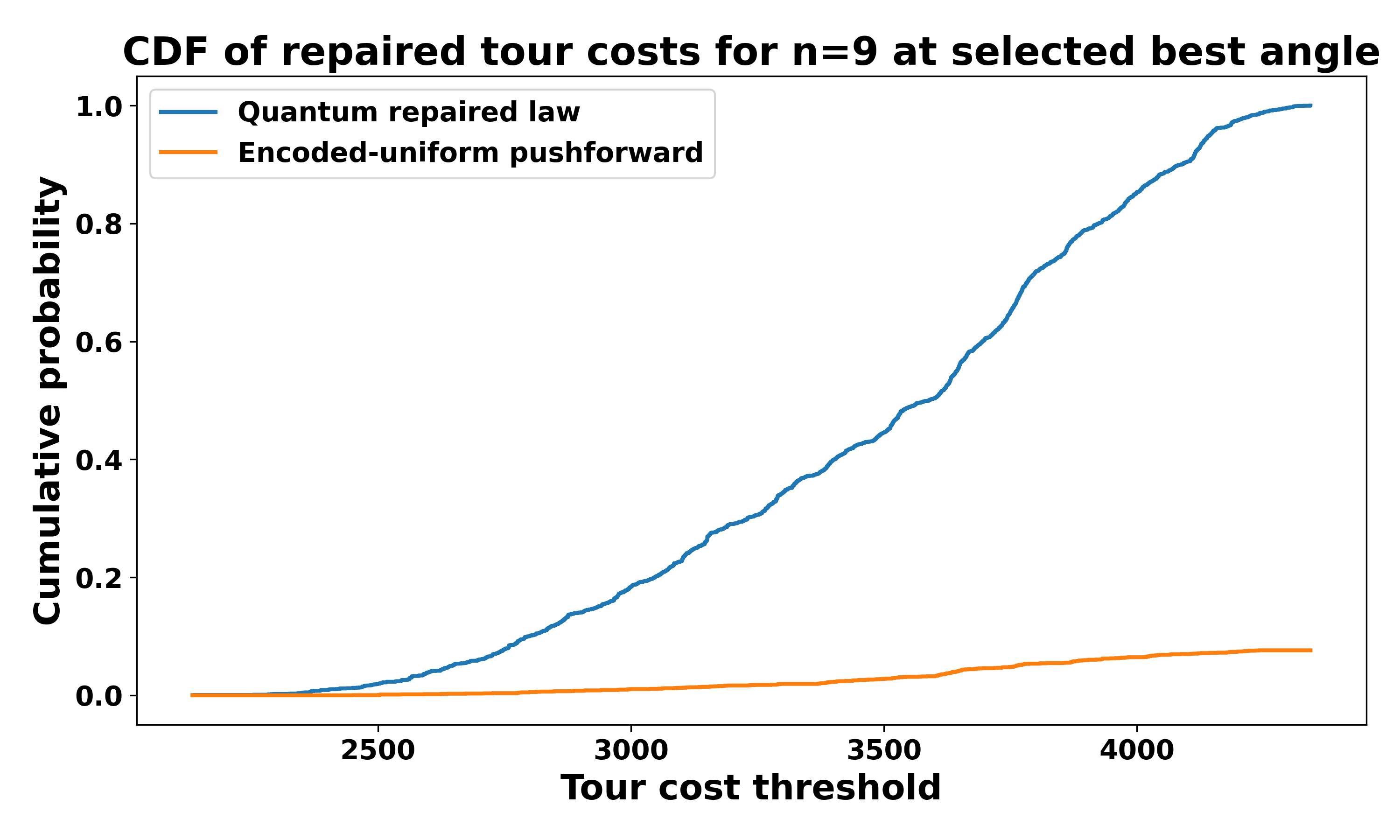}
        \caption{\(n=9\)}
        \label{fig:cdf-n9}
    \end{subfigure}
    \hfill
    \begin{subfigure}[t]{0.48\textwidth}
        \centering
        \includegraphics[width=\textwidth]{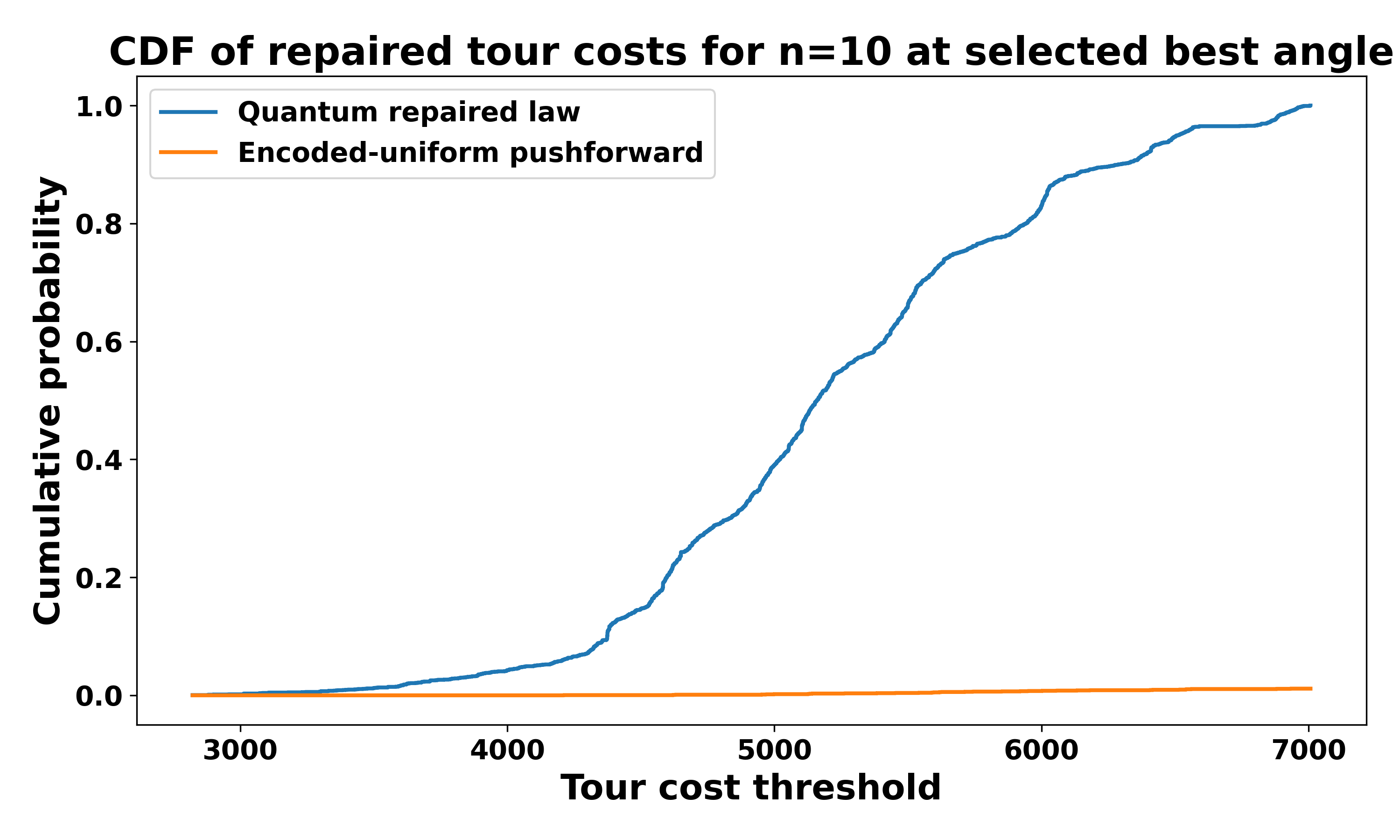}
        \caption{\(n=10\)}
        \label{fig:cdf-n10}
    \end{subfigure}

    \caption{CDFs of repaired tour costs at the selected best angle
    for \(n=7,8,9,10\). The blue curves represent the repaired quantum
    laws and the orange curves represent the encoded-uniform pushforward
    baselines obtained with the same deterministic Hungarian repair map.
    The quantum laws accumulate greater probability on lower-cost
    feasible tours over substantial portions of the plotted ranges,
    with a widening separation across the tested instances.}
    \label{fig:repaired-cdfs}
\end{figure*}

For a cost threshold \(c\), define the repaired cost cumulative
distribution function by
\begin{equation}
    F_R(c)
    :=(R_\#p)\!\left(\{\pi\in S_n:C(\pi)\leq c\}\right).
\end{equation}
At the known optimal tour cost \(C^\star:=C(\pi^\star)\), this is the
probability of recovering a tour of optimal cost, including all optimal
permutations. Figure~\ref{fig:repaired-cdfs} compares the empirical
quantum repaired CDF with the shot-matched encoded-uniform pushforward
baseline.

The repaired quantum CDF lies above the encoded-uniform 
baseline across substantial portions of the low-cost range in every
instance. The separation becomes more pronounced as \(n\) increases.
For \(n=10\), the baseline remains near zero through most of the plotted low-cost region, while the quantum repaired law accumulates substantial probability over low-cost tours. Because both laws are
repaired by the same map, these differences are differences in the raw probability assigned to the preimages of the corresponding sets.

\begin{figure*}[t]
    \centering
    \begin{minipage}[t]{0.49\textwidth}
        \centering
        \includegraphics[width=\linewidth]{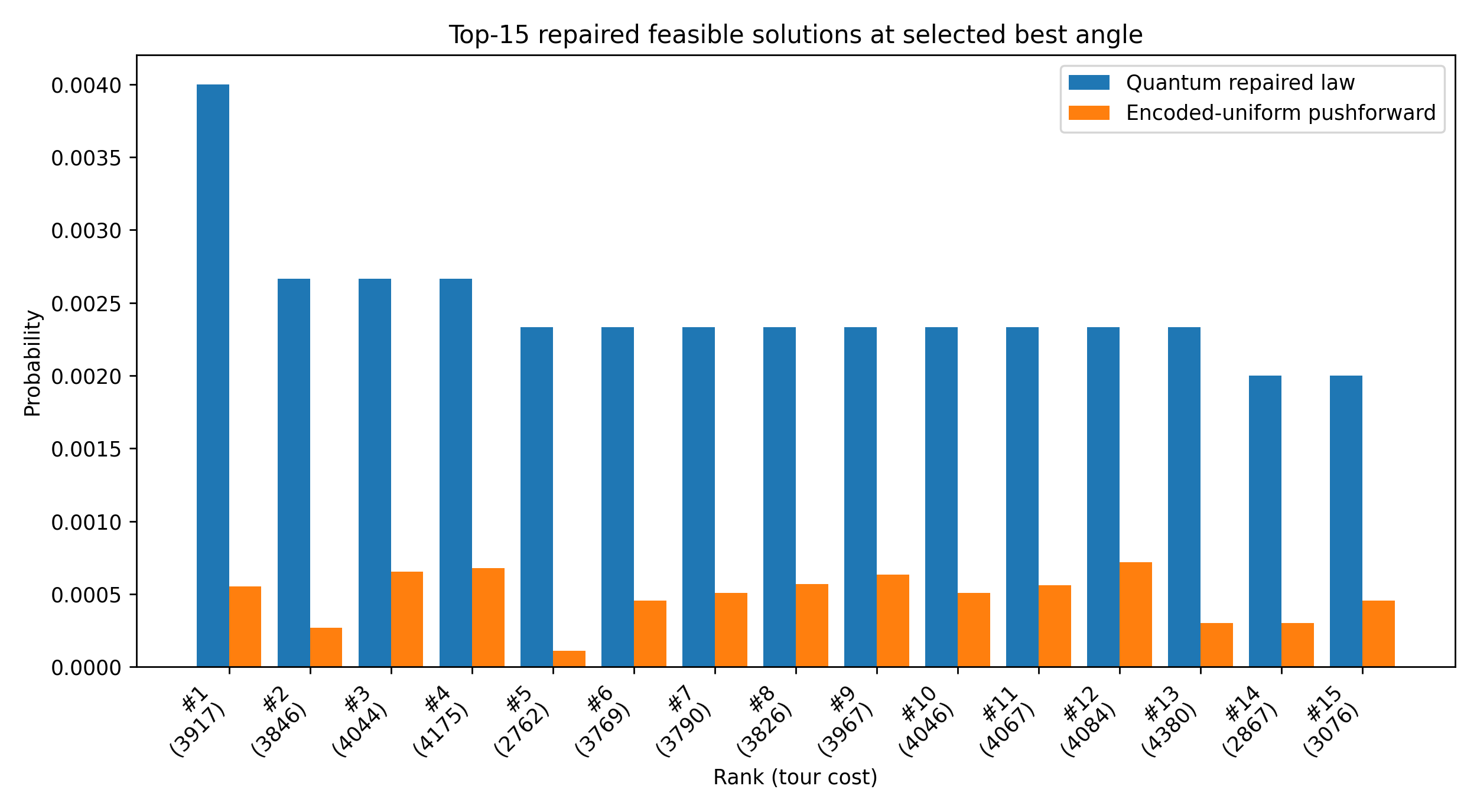}
        \small (a) dj8
    \end{minipage}
    \hfill
    \begin{minipage}[t]{0.49\textwidth}
        \centering
        \includegraphics[width=\linewidth]{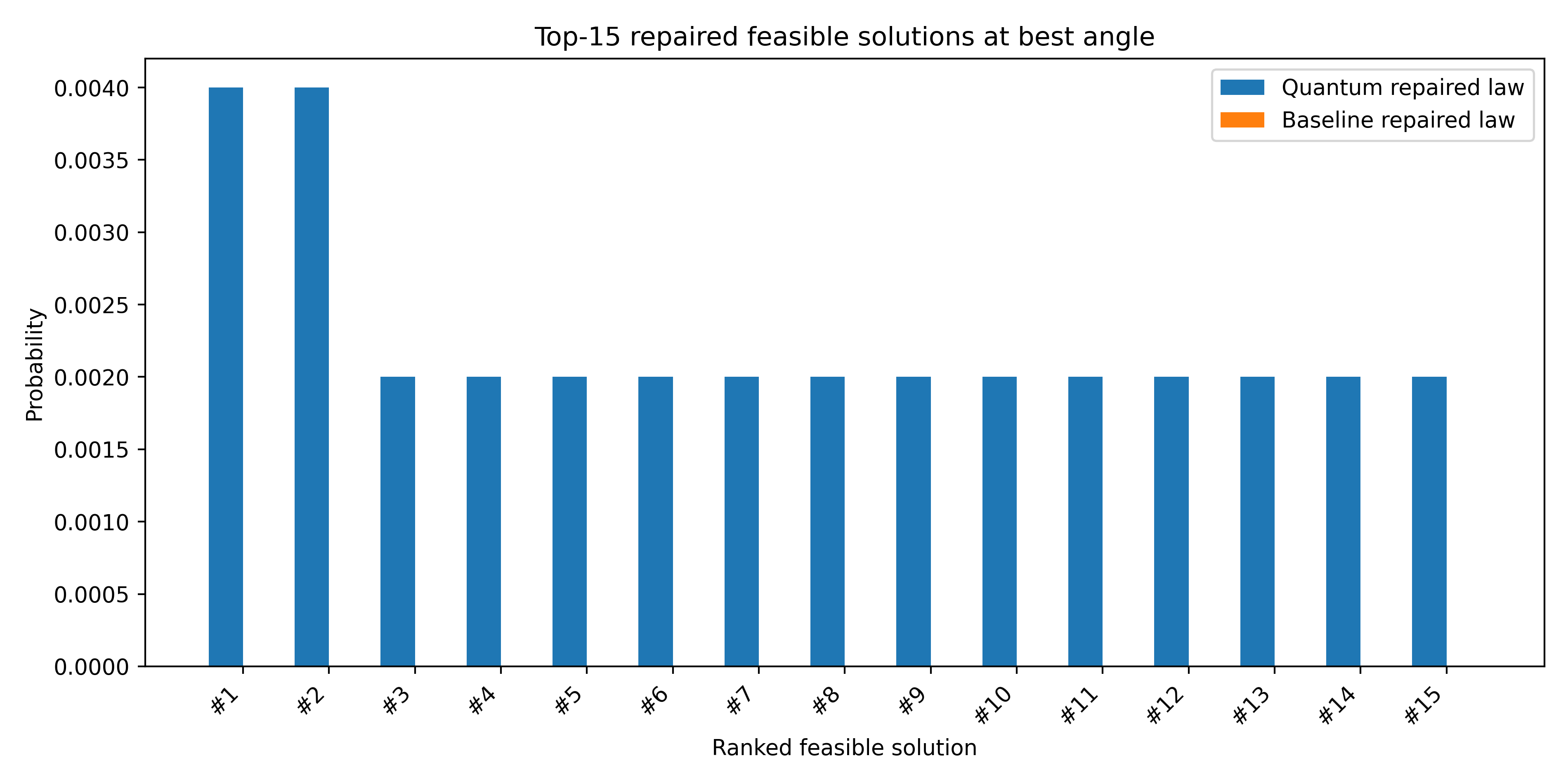}
        \small (b) dj9
    \end{minipage}
    \caption{Solution-level comparison of the repaired feasible sampling  laws for the dj8 and dj9 TSP instances at a selected angle.  Each panel displays the 15 best feasible solutions ranked by  empirical quantum repaired probability. The blue bars show the     quantum pushforward \(R_\#p\), and the orange bars show the
    encoded-uniform pushforward \(R_\#u\) under the same deterministic
    Hungarian repair map. The quantum law assigns greater probability
    to each displayed solution, with the baseline strongly suppressed
    for dj9 at the displayed resolution. These plots complement the
    cost-threshold CDFs in Fig.~\ref{fig:repaired-cdfs}.}
    \label{fig:repaired-top15-comparison}
\end{figure*}

A complementary solution-level comparison is shown in
Fig.~\ref{fig:repaired-top15-comparison}. For dj8 and dj9, we display the
15 leading feasible solutions ranked by their empirical quantum repaired
probabilities and evaluate the encoded-uniform pushforward probabilities
of those same solutions. The quantum law assigns greater probability to
each displayed solution, with a particularly strong separation for dj9.
Thus, the cumulative preference for lower-cost tours seen in the CDFs is
accompanied by a solution-by-solution concentration of quantum probability
on particular repair preimages.

The deterministic tie-breaking convention affects the sizes of the
repair basins and hence the encoded-uniform pushforward itself. To
quantify this effect, we enumerated all \(6^6\) row-one-hot words at
\(n=6\). Lexicographically first maximum-overlap repair and the
matching solution returned by SciPy 1.17.0
\cite{virtanen2020scipy}, using a zero--one mismatch matrix, assign
different permutations to \(35.46\%\) of these words. Their respective encoded-uniform repaired laws have total-variation distances \(0.395\)
and \(0.208\) from the uniform law on permutations. The stochastic
symmetry-controlled repair of Appendix~\ref{app:fair-repair} has
zero total-variation distance. These exact small-size calculations
illustrate the role of a fixed tie rule in determining the classical
baseline. The quantum and encoded-uniform comparisons above use the
same deterministic repair implementation for each instance.

\begin{figure*}[t]
    \centering
    \begin{subfigure}[t]{0.48\textwidth}
        \centering
        \includegraphics[width=\textwidth]{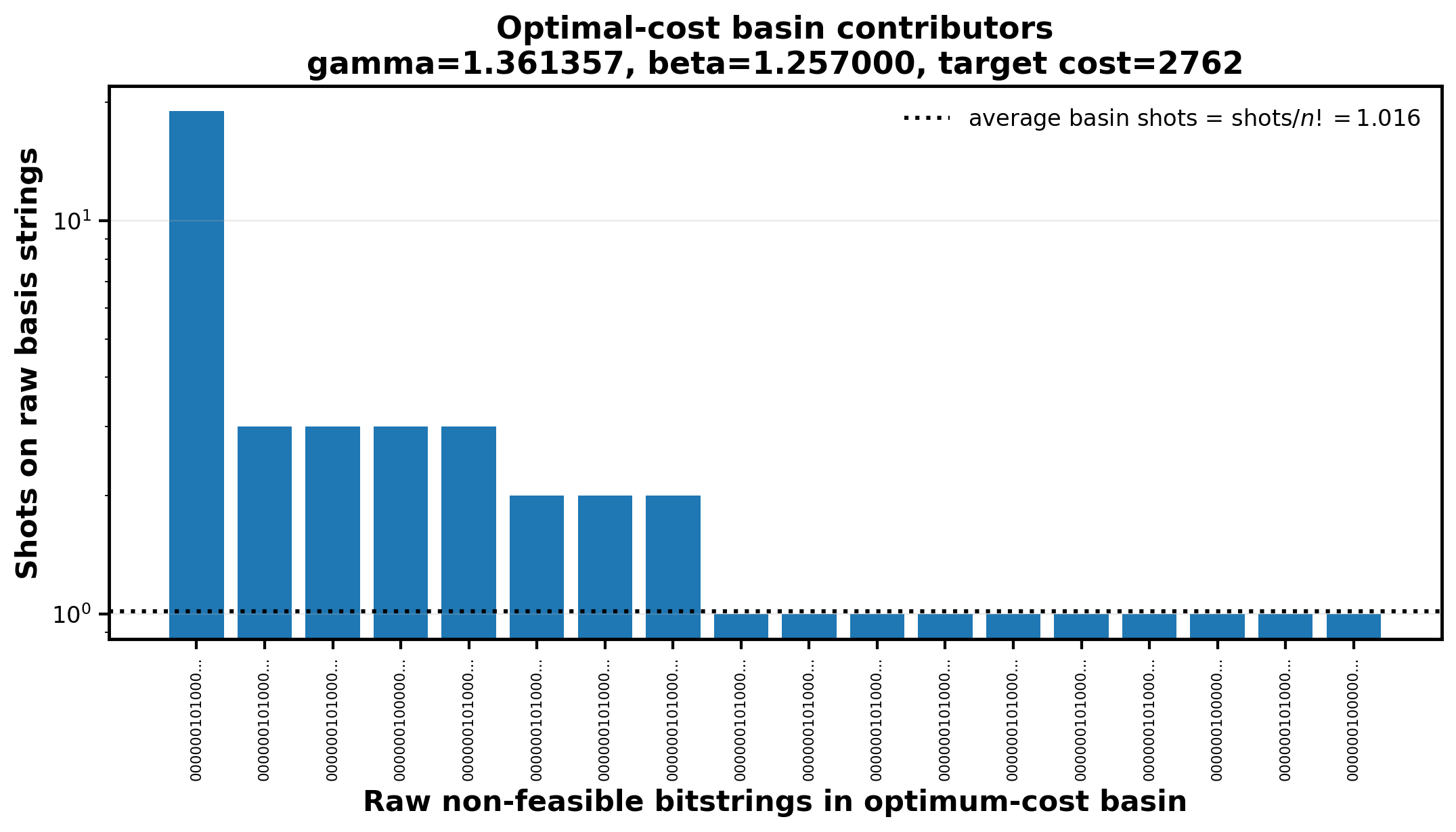}
        \caption{Optimal-cost contributors at
        \((\gamma,\beta)=(1.361357,\,1.257000)\).}
        \label{fig:basin-opt-cost-1361-1257}
    \end{subfigure}
    \hfill
    \begin{subfigure}[t]{0.48\textwidth}
        \centering
        \includegraphics[width=\textwidth]{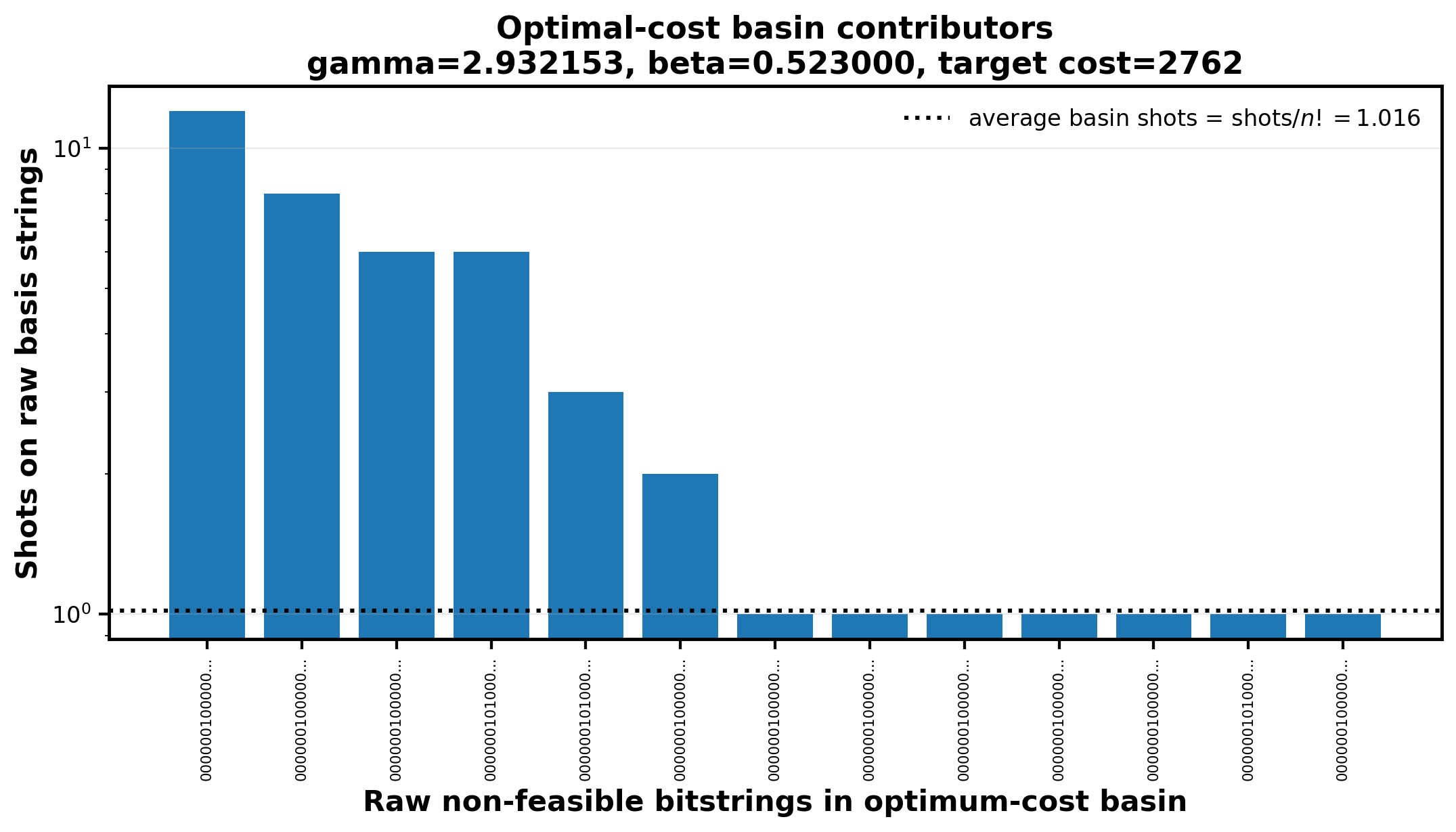}
        \caption{Optimal-cost contributors at
        \((\gamma,\beta)=(2.932153,\,0.523000)\).}
        \label{fig:basin-opt-cost-2932-0523}
    \end{subfigure}
    \caption{Raw contributors to the optimal-cost repair
    preimage at two CE--QAOA angle pairs. Each bar records the number
    of shots assigned to a raw encoded basis string that repairs to a
    tour of optimal cost \(2762\). The dotted line is \(S/n!\), the
    expected shot count of an entire average permutation basin under
    encoded-uniform sampling. Several individual contributors exceed
    this reference in both panels.}
    \label{fig:optimal-cost-basin-contributors}
\end{figure*}

Figure~\ref{fig:optimal-cost-basin-contributors} resolves the quantum
concentration further into individual raw configurations. It shows
infeasible encoded strings that repair to tours of optimal cost
\(2762\) at two CE--QAOA angle pairs. Several individual strings
receive more shots than the number \(S/n!\) expected for an entire
average permutation basin under encoded-uniform sampling. These
counts identify raw, infeasible contributors to the successful
repaired event and show the strength of their quantum weighting.

The combinatorial and dynamical contributions can be distinguished
using the basin factors introduced in
\cref{thm:log-depth-threshold,subsec:discussion-volume-concentration}.
For a fixed permutation \(\pi\), the encoded-uniform reference assigns
probability \(\kappa_\pi/n!\), while the quantum law assigns
\((R_\#p)(\pi)=(\kappa_\pi/n!)\chi_\pi\). The CDFs aggregate these
probabilities over cost-defined feasible sets, the dj8/dj9 bar plots
resolve individual feasible outputs, and the contributor histograms
identify raw configurations supplying quantum mass to optimal-cost
preimages. Together, these comparisons demonstrate the role of both
repair-basin volume and instance-dependent quantum concentration in
the repaired output law.

%======================================================================
\subsection{Harvesting Quantum Bias on Noisy Superconducting Hardware}
\label{sec:experiment}
%======================================================================

We analyze the hardware measurements reported in Ref.~\cite{Onah2026GeometryInformed} using
the repaired sampling laws considered here. The presentation focuses
on the probability assigned to preimages of good tours under the
specified repair map. The ideal CE--QAOA evolution produces outcomes in the
row-one-hot encoded space \(\mathcal X\). Hardware errors can produce
bit strings outside this sector, so the relevant raw measurement space is
the ambient cube
\[
    \Omega=\{0,1\}^{n^2},
    \qquad \mathcal X\subseteq\Omega.
\]
Hungarian maximum-overlap repair extends to a deterministic map
\(R:\Omega\to S_n\), taking any measured binary assignment matrix to
a maximum-overlap feasible permutation under the fixed tie rule. The
noisy hardware law \(p^{\mathrm{hw}}\) therefore induces
\begin{equation}
    p_R^{\mathrm{hw}}:=R_\#p^{\mathrm{hw}},
\end{equation}
with
\begin{equation}
\label{eq:hardware-pushforward-law}
    p_R^{\mathrm{hw}}(\pi)
    =\sum_{z\in\Omega:\,R(z)=\pi}p^{\mathrm{hw}}(z),
    \qquad \pi\in S_n.
\end{equation}

For \(S\) measured samples, Hungarian repair\cite{Kuhn1955Hungarian} costs \(O(Sn^3)\) in
total and produces a feasible permutation from every raw outcome.
The empirical repaired law is
\begin{equation}
\label{eq:empirical-hardware-pushforward}
    \widehat p_R^{\mathrm{hw}}(\pi)
    =\frac{1}{S}\sum_{s=1}^{S}
    \mathbf 1\!\left\{R\!\left(Z^{(s)}\right)=\pi\right\},
    \qquad Z^{(s)}\sim p^{\mathrm{hw}}.
\end{equation}
The hardware experiment measures how much raw probability is transported into the repair preimages of low-cost feasible tours. As discussed in \cref{subsec:noise-leakage-basin-breakdown} and in  \cref{eq:repaired-success-decomposition}, in the presence of encoded-subspace leakage, decompose the noisy
measurement law as
\begin{equation}
    p^{\mathrm{hw}}
    =(1-\eta)\,p^{X}+\eta\,p^{\mathrm{out}},
\end{equation}
where \(p^{X}\) and \(p^{\mathrm{out}}\) are supported on
\(\mathcal X\) and \(\Omega\setminus\mathcal X\), respectively.
The repaired hardware law then satisfies
\begin{equation}
\label{eq:hardware-sector-pushforward}
    R_\#p^{\mathrm{hw}}
    =(1-\eta)\,R_\#p^{X}
     +\eta\,R_\#p^{\mathrm{out}}.
\end{equation}
The first term represents repaired contributions from outcomes
remaining in the encoded space; the second collects repaired
contributions from leaked outcomes. Their sum is the complete
repaired output distribution measured in the experiment.

We use single-layer CE--QAOA circuits on the 127-qubit IBM Eagle-r3
processor \texttt{ibm\_brisbane}, with \textsc{QOptlib} TSP subset
instances containing up to \(n=10\) cities, or \(100\) binary assignment
variables~\cite{Osaba2024Qoptlib}. The shallow circuit construction
limits coherent depth while the repair operation processes every
measured bit string, including outcomes that violate the row-one-hot
encoding.

\subsubsection{Benchmark instances and transferred single-layer parameters}
\label{subsec:hardware-benchmarks}

Seven instances, \(n=4,\ldots,10\), require \(n^2=16\) through \(100\)
qubits for their binary assignment encodings. We use the \(p=1\)
parameter pairs obtained in previous generic-QAOA optimizations and
transfer them to CE--QAOA without further hardware parameter
optimization~\cite{Onah2025qprobe}. The transferred parameters are
reported in Table~\ref{tab:qaoa:p1-warmstarts}.

\begin{table}[ht]
    \centering
    \begin{tabular}{lccc}
        \toprule
        Instance & \(\gamma^{\mathrm{gen}}\)
                 & \(\beta^{\mathrm{gen}}\)
                 & Best generic-QAOA energy \\
        \midrule
        wi4  & 0.233 & 2.025 & \(-4.36\times10^{4}\) \\
        wi5  & 1.307 & 2.387 & \(-6.11\times10^{4}\) \\
        wi6  & 0.748 & 0.741 & \(-4.14\times10^{4}\) \\
        wi7  & 2.172 & 0.577 & \(-7.66\times10^{4}\) \\
        dj8  & 0.239 & 0.460 & \(-3.18\times10^{4}\) \\
        dj9  & 0.002 & 2.507 & \(-1.72\times10^{6}\) \\
        dj10 & 1.835 & 2.399 & \(-1.68\times10^{4}\) \\
        \bottomrule
    \end{tabular}
    \caption{Generic-QAOA parameters obtained at \(p=1\) and transferred
    to the CE--QAOA hardware circuits. No further parameter
    optimization was performed on the processor.}
    \label{tab:qaoa:p1-warmstarts}
\end{table}

All circuits were implemented in Python and transpiled with
\textsc{Qiskit~1.1.2}~\cite{Qiskit2023}, using device-aware coupling maps
and the available native two-qubit gate basis. For each instance, we
executed a single CE--QAOA layer with
\begin{equation}
    (\gamma,\beta)
    =\bigl(\gamma^{\mathrm{gen}},\beta^{\mathrm{gen}}\bigr).
\end{equation}
Measurement-error mitigation and dynamical decoupling were enabled.
For each measured outcome \(Z^{(s)}\), we recorded its repaired image
\begin{equation}
    Y^{(s)}:=R\!\left(Z^{(s)}\right)\in S_n.
\end{equation}
The empirical raw and repaired feasible fractions are
\begin{equation}
\label{eq:hardware-raw-feasible-fraction}
    \widehat P_{\mathrm{raw}}(F)
    =\frac{1}{S}\sum_{s=1}^{S}
      \mathbf 1\!\left\{Z^{(s)}\in S_n\right\},
\end{equation}
and
\begin{equation}
\label{eq:hardware-repaired-feasible-fraction}
    \widehat P_{\mathrm{rep}}(F)
    =\frac{1}{S}\sum_{s=1}^{S}
      \mathbf 1\!\left\{R\!\left(Z^{(s)}\right)\in S_n\right\}
    =1.
\end{equation}

The best repaired tour cost observed in a run is
\begin{equation}
\label{eq:hardware-best-repaired-cost}
    C_{\min}^{\mathrm{rep}}
    =\min_{1\leq s\leq S}C\!\left(R\!\left(Z^{(s)}\right)\right).
\end{equation}
For a cost threshold \(\tau\), let
\(
    G_\tau:=\{\pi\in S_n:C(\pi)\leq\tau\}.
\)
The empirical repaired probability of meeting the threshold is
\begin{equation}
\label{eq:hardware-good-basin-mass}
    \widehat P_{\mathrm{rep}}(G_\tau)
    =\widehat p^{\mathrm{hw}}\!\left(R^{-1}(G_\tau)\right).
\end{equation}
Thus, each observed low-cost repaired tour identifies a raw
measurement in a preimage of the corresponding feasible output.

%======================================================================
\subsection{IBM Brisbane: 127-qubit Eagle-r3 processor}
\label{ssec:exp-brisbane}
%======================================================================

Table~\ref{tab:snap} summarizes the device characteristics recorded
for the hardware run. The median two-qubit ECR error was
\(6.99\times10^{-3}\), with a best value of \(2.67\times10^{-3}\), while the median readout error was \(1.61\times10^{-2}\). The larger circuits undergo noisy encoded evolution and subsequent
measurement on the full binary outcome space. In the language of
\cref{subsec:noise-leakage-basin-breakdown}, their repaired law may
inherit both probability distorted within \(\mathcal X\) and
probability leaked into \(\Omega\setminus\mathcal X\). The experiment
therefore probes the complete noisy pushforward \(R_\#p^{\mathrm{hw}}\).

\begin{table}[ht]
    \centering
    \begin{tabular}{@{}lcc@{}}
        \toprule
        \textbf{Metric} & \textbf{Median value} & \textbf{Best} \\
        \midrule
        Qubits & 127 & -- \\
        Two-qubit ECR error \(\epsilon_{2Q}\)
            & \(6.99\times10^{-3}\) & \(2.67\times10^{-3}\) \\
        Readout error
            & \(1.61\times10^{-2}\) & \(5.1\times10^{-3}\) \\
        \bottomrule
    \end{tabular}
    \caption{Device characteristics of \texttt{ibm\_brisbane} during
    the reported experiments.}
    \label{tab:snap}
\end{table}

Table~\ref{tab:qoptlib-comparison} compares the best repaired candidates
against the reference tour costs distributed with
\textsc{QOptlib}~\cite{Osaba2024Qoptlib}. The hardware results
reproduce the published optimal costs for the smaller \texttt{wi}
instances and improve on the distributed reference costs for the
larger \texttt{dj} subsets. The repaired values reported here are
known optimal costs, as previously established in
Refs.~\cite{onahce,Onah2026GeometryInformed}. For the largest instance, the hardware experiment used \(10{,}000\)
shots, while the noiseless pushforward simulation in
\cref{subsec:noiseless-analysis} recovered an optimum in \(1{,}000\)
shots. This difference is consistent with the noise-related loss described in \cref{subsec:noise-leakage-basin-breakdown}. The original noiseless CE--QAOA simulations without pushforward in
Ref.~\cite{onahce} used \(10^6\) shots. However, these figures describe the sampling budgets of the respective experiments under different parameterizations. 

% \sh{These shot counts describe different experimental workflows and do not establish a reduction in sampling cost. Such a comparison requires independently estimated success
% probabilities for the same event, together with the costs of parameter selection, circuit execution,
% and classical processing.}

\begin{table}[H]
    \centering
    \caption{Best repaired tours obtained from \texttt{ibm\_brisbane}
    compared with the reference values distributed by
    \textsc{QOptlib}~\cite{Osaba2024Qoptlib}. A positive improvement
    indicates a shorter repaired tour than the reference.}
    \label{tab:qoptlib-comparison}
    \begin{tabular}{lrrrr}
        \toprule
        Instance & QOPTLib cost & Repaired hardware cost
                 & \(\Delta\) & Improvement [\%] \\
        \midrule
        wi4  & 6\,700 & 6\,700 & 0 & 0.00 \\
        wi5  & 6\,786 & 6\,786 & 0 & 0.00 \\
        wi6  & 9\,815 & 9\,815 & 0 & 0.00 \\
        wi7  & 7\,245 & 7\,245 & 0 & 0.00 \\
        dj8  & 2\,794 & 2\,762 & \(-32\) & \textbf{1.15} \\
        dj9  & 2\,438 & 2\,134 & \(-304\) & \textbf{12.5} \\
        dj10 & 3\,155 & 2\,822 & \(-333\) & \textbf{10.55} \\
        \bottomrule
    \end{tabular}
\end{table}

The \textsc{QOptlib} figures in Table~\ref{tab:qoptlib-comparison}
are the supplied benchmark reference values. Osaba \emph{et al.}
reported tour costs \(2762\), \(2134\), and \(2822\) on the same dj8, dj9, and dj10 instances in ten independent runs of Quantagonia's hybrid solver~\cite{osaba2024hybrid}. These match the corresponding repaired hardware costs above. The values were also certified as optimal using
mixed-integer programming solvers in earlier work
by two of us~\cite{onahce}.

\section{Discussion}

The pushforward map changes the event counted as
success from sampling a particular feasible
configuration to sampling any raw configuration
in its repair basin. At the combinatorial level,
this enlarges the successful event by the volume
of the repair preimage. At the dynamical level,
the enlarged event reduces the amount of coherent
shell-transfer amplification required to reach
a fixed success-probability at a fixed depth. The
interplay of these two effects leads to further
consequences which we now discuss.

\subsection{From pointwise amplification to basin amplification}
\label{subsec:discussion-pointwise-to-basin}

The shell-recursion formalism of
Ref.~\cite{onahseperating} characterizes amplitude
transport to a target configuration in the
encoded product space. For a pointwise successful
event, the shell-transfer factor must compensate
the normalization associated with a single basis
state. The corresponding depth condition scales
logarithmically with the local alphabet size,
taking the form
\(
p\gtrsim\frac{1}{2}\log_2 n
\)
in the normalization of Ref.~\cite{onahseperating}. Pushforward repair enlarges the successful event
from a fixed optimal permutation $\pi^\star$ to
its entire repair basin, \( A_\star\) where every configuration in $A_\star$ repairs to an optimal solution.

The repair-basin volume enters directly into the
amplification requirement. Since the repair
basins partition the encoded space
$\mathcal X=[n]^n$, their average cardinality
is $n^n/n!$. Using the relative
optimal-basin volume introduced in
\cref{thm:log-depth-threshold},
\[
\kappa_\star
=
\frac{n!|A_\star|}{n^n},
\]
the repaired-success bound becomes
\[
P^{\mathrm{rep}}_{\mathrm{CE},p}(\pi^\star)
\geq
\frac{\kappa_\star}{n!}
\cos^2\!\left(\frac{\Theta_p}{2}\right)
\prod_{\ell=1}^{p}
q_n(\beta_\ell)^{2n}.
\]
At the average basin scale
$\kappa_\star=1$, the combinatorial factor
$n^n/n!$ converts the pointwise normalization
$n^{-n}$ into the factorial scale $1/n!$.
The condition for absolute transfer to
compensate the optimal-basin normalization is
\(
q_{\max}(n)^{2np}
\geq
\frac{n!}{\kappa_\star}.
\)
Consequently, the compensation depth is
\begin{equation}
\label{eq:discussion-general-pmin}
p_{\mathrm{abs}}(n,\kappa_\star)
=
\left\lceil
\frac{
\log n!-\log\kappa_\star
}{
2n\log(3-4/n)
}
\right\rceil.
\end{equation}

The factorial term $\log n!$
determines the average-basin compensation
threshold, while $-\log\kappa_\star$ in Equation~\eqref{eq:discussion-general-pmin}
accounts for the volume of the
optimal repair basin. For
$\kappa_\star>1$, the larger preimage
reduces the required circuit depth. The resulting reduction is especially
relevant at finite problem sizes because
circuit depth is integer valued. A change
in the continuous compensation threshold
can remove an entire coherent layer.
For the benchmark sizes $n=7,8,9,10$,
the absolute-transfer condition already
admits $p=1$ at the average basin scale
$\kappa_\star=1$. Thus, pushforward repair contributes to
sampling efficiency through the aggregation
of probability over a larger successful event
and the corresponding reduction in the
absolute-transfer amplification requirement.
% Under the phase-alignment conditions of
% Corollary~\ref{cor:uniform-phase-aligned-repair},
% these contributions enter the finite-shot
% guarantees of
% \cref{thm:log-depth-threshold}.

\subsection{Single-layer CE--QAOA and reduced depth as a phase-control resource}
\label{subsec:discussion-depth-phase-control}

The one-layer regime is particularly relevant
to the numerical and hardware experiments
considered in this work. At shallow depth,
the quantum circuit performs fewer coherent
operations, while the repair map aggregates
the resulting probability over the preimages of feasible solutions. The depth reduction also has a dynamical
consequence. Every additional CE--QAOA
layer introduces another cost-phase
operation and mixer action, enlarging the
family of interfering paths contributing
to a measured configuration. Reducing
the required amplification depth
therefore reduces the number of successive
operations involved in organizing
these contributions. Pushforward repair changes this resource
requirement through the size of the
successful event. A larger repair basin
requires less absolute amplitude transport
to reach the same repaired-success
threshold. Consequently, repair geometry
can reduce both the coherent depth
associated with amplification and the
complexity of the corresponding
phase-control task.

To quantify the finite-size effect,
consider a single CE--QAOA layer at
the transport-maximizing angle
\(
\beta^\star=\frac{\pi(n-1)}{n},
q_n(\beta^\star)=3-\frac{4}{n}.
\)
Write the optimal repair-basin size as
\[
|A_\star|
=
f_n\frac{n^n}{n!},
\qquad
f_n=\kappa_\star.
\]
For a polynomial shot budget $N=n^k$,
define the depth-one pushforward
transport budget
\begin{equation}
\label{eq:p1-pushforward-transport-budget}
\mathcal B_1(n;f_n,k)
:=
\frac{f_n n^k}{n!}
\left(3-\frac{4}{n}\right)^{2n}.
\end{equation}
The condition $B_1\geq 1$ identifies an absolute-transfer budget
crossover. For a target failure probability $\epsilon$, the
finite-shot guarantee instead requires
$$ B_1c_1 \geq \log(1/\epsilon),
    \qquad
    c_1 := \cos^2(\Theta_1/2),$$
together with the common-arc hypothesis. 

% This combines the optimal
% repair-basin volume, the maximal
% single-layer absolute transfer factor,
% and the available sampling budget.
% The condition
% \(
% \mathcal B_1(n;f_n,k)\geq1
% \)
% identifies the finite-size regime in
% which their combined absolute-transfer
% budget reaches unity. 

\begin{figure}[t]
    \centering
    \includegraphics[width=\linewidth]
    {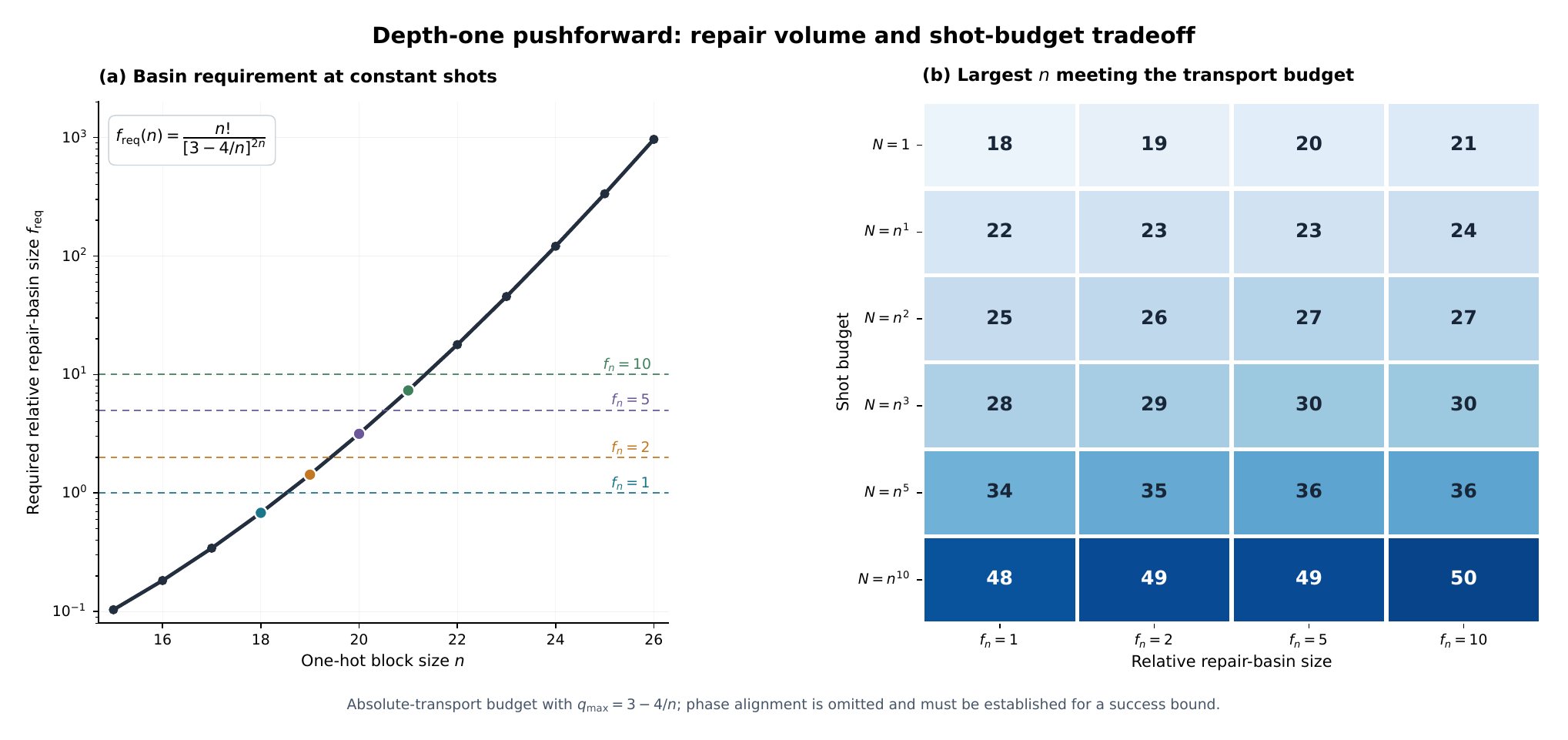}
    \caption{
    Depth-one pushforward transport compensation
    at the maximal absolute-transfer angle
    $\beta^\star=\pi(n-1)/n$.
    (a) Required relative optimal-basin volume
    $f_{\mathrm{req}}(n)
    =n!/(3-4/n)^{2n}$
    for unit-shot absolute-transfer compensation.
    Colored markers indicate the largest
    problem size satisfying the criterion for
    $f_n\in\{1,2,5,10\}$.
    (b) Largest problem size for which the
    depth-one transport budget reaches unity
    with $N=n^k$ shots.
    Both panels quantify absolute-transfer
    compensation; the finite-shot
    repaired-success guarantee additionally
    incorporates the phase-alignment factor.
    }
    \label{fig:p1-pushforward-budget}
\end{figure}

Figure~\ref{fig:p1-pushforward-budget}
shows how the relative optimal-basin
volume and polynomial sampling budgets
affect this threshold. Panel (a)
identifies the basin enhancement required
for single-layer compensation at a fixed
unit-shot budget, while panel (b)
shows the corresponding increase in
the largest problem size satisfying the absolute transfer criterion  when the shot budget scales as $N=n^k$.
The corresponding probability guarantees
follow under the phase-alignment
conditions in
Corollary~\ref{cor:uniform-phase-aligned-repair}.
The derivation and finite-size analysis
are given in
Appendix~\ref{app:basin-shot-depth-tradeoff}. The depth-one analysis demonstrates how repair-basin enlargement and classical sampling resources can support shallower coherent circuits. This is relevant to noisy devices, where each additional quantum layer introduces further gate operations, implementation errors, and opportunities for leakage from the encoded subspace.

An important observation is that the average repair-basin size $n^n/n!$
already captures much of the potential resource enhancement. At $N=n^3$, increasing the relative basin volume fivefold extends the single-layer
threshold only from $n=28$ to $n=30$. Even increasing the shot budget to $N=n^5$
extends the threshold only to $n=34$ at the average-basin scale. These diminishing returns highlight the limitations of classical sampling resources and further basin enlargement, motivating additional coherent quantum resources to access larger problem sizes. The persistence of quantum-classical separation in this type of hybrid workflow was briefly discussed in Ref. \cite{Onah2026GeometryInformed}.

\subsection{Combinatorial basin volume and quantum basin concentration}
\label{subsec:discussion-volume-concentration}

For a fixed optimal permutation $\pi^\star$, its basin
$A_\star=R^{-1}(\pi^\star)$ occupies
the fraction
\(
\frac{|A_\star|}{|\mathcal X|}
=
\frac{\kappa_\star}{n!}
\)
and the probability of recovering $\pi^\star$ when the
encoded-uniform sampling is followed by Hungarian repair is:
\(
(R_\#u)(\pi^\star)
=
\frac{|A_\star|}{|\mathcal X|}
=
\frac{\kappa_\star}{n!}.
\)
For CE--QAOA, the repaired probability is
\(
(R_\#p)(\pi^\star)
=
\sum_{x\in A_\star}p(x).
\)
The difference between the repaired
quantum law and the uniform pushforward is
\begin{equation}
\label{eq:discussion-excess-basin-mass}
(R_\#p)(\pi^\star)
-
(R_\#u)(\pi^\star)
=
\sum_{x\in A_\star}
\left(
p(x)-\frac{1}{|\mathcal X|}
\right).
\end{equation}
This quantity measures the excess
quantum probability assigned to the
optimal repair basin. The distinction can also be expressed
through the basin concentration factor
\[
\chi_\star
:=
\frac{(R_\#p)(\pi^\star)}
     {(R_\#u)(\pi^\star)}
=
\frac{n!}{\kappa_\star}
\sum_{x\in A_\star}p(x).
\]
The repaired optimum probability then
admits the exact factorization
\begin{equation}
(R_\#p)(\pi^\star)
=
\frac{\kappa_\star}{n!}\chi_\star.
\end{equation}
Here $\kappa_\star$ quantifies the
combinatorial volume supplied by repair,
while $\chi_\star$ quantifies the relative
concentration supplied by the quantum
sampling law. Values $\chi_\star>1$
indicate that CE--QAOA assigns greater
probability to the optimal repair basin
than encoded-uniform sampling. The numerical results exhibit this
distinction at two complementary levels. The repaired cost CDFs and the solution-level comparisons in
Figs.~\ref{fig:repaired-cdfs} and
\ref{fig:repaired-top15-comparison}
show how quantum probability is
concentrated on low-cost feasible tours
and their respective repair preimages.
Because both sampling laws are defined
on the same row-one-hot encoded space
and use the same repair map, these comparisons  measure
differences in their preimage probabilities.

The contributor histograms in \cref{fig:optimal-cost-basin-contributors} resolve this concentration further
at the level of encoded configurations. Several raw
strings belonging to an optimum-cost repair preimage receive quantum counts exceeding the average-basin benchmark $\mathrm{shots}/n!$.
These contributors demonstrate how instance-dependent quantum dynamics can populate useful repair preimages
far above the uniform reference scale.

\subsection{Implications for near-term algorithm design}
\label{subsec:discussion-hybrid-design}

The preceding analysis suggests that the repair map determines the partition of the encoded configuration space into feasible-solution preimages,
\begin{equation}
    \mathcal X
    =
    \bigsqcup_{y\in F}R^{-1}(y),
\end{equation}
while the quantum circuit determines how probability
is distributed across this partition. The two operations
therefore contribute complementary resources to the
hybrid algorithm. For a target set $G\subseteq F$ of high-quality
feasible solutions, the relevant successful event
is the complete preimage $R^{-1}(G)$. This suggests designing repair maps  together with quantum circuits that concentrate probability on those preimages at shallow depth. Such a design objective is naturally expressed through the repaired success probability
\begin{equation}
    P^{\mathrm{rep}}_{\mathrm{CE},p}(G)
    =
    p\!\left(R^{-1}(G)\right),
\end{equation}
or the repaired expected objective
\begin{equation}
    \mathbb E_p[C(R(Z))].
\end{equation}
Both quantities incorporate the complete
quantum--classical sampling pipeline and can
be evaluated directly from repaired measurement
outcomes. They also provide natural objectives
for selecting circuit parameters and comparing
candidate repair strategies.

The separation between basin volume and quantum
concentration further clarifies the corresponding
algorithmic resources. As established in
\cref{subsec:discussion-volume-concentration},
with the factorization
\[
    (R_\#p)(\pi^\star)
    =
    \frac{\kappa_\star}{n!}\chi_\star,
\]
the factor $\kappa_\star$ is determined by the
repair geometry, while $\chi_\star$ measures
the quantum concentration relative to the
encoded-uniform reference. Co-design
can therefore investigate both the geometry
of useful repair preimages and the dynamics
that populate them.

% The depth reduction supplied by basin aggregation
% is particularly relevant to noisy quantum devices.
% Shallower coherent circuits require fewer
% successive quantum operations and can reduce
% the accumulated implementation error.
% As discussed in
% \cref{subsec:noise-leakage-basin-breakdown},
% noise can degrade repaired sampling through
% leakage from the encoded subspace and loss
% of concentration within that subspace.
% Reducing the absolute-transfer requirement
% through repair-basin enlargement provides
% an additional route to limiting the coherent
% resources required for useful sampling.

The computational accessibility of favorable
sampling distributions is also an important
consideration. For MaxCut, Ref. ~\cite{hadfield2026complexity} established worst-case complexity barriers to uniformly
efficient state preparation achieving a fixed
fraction of the optimal gain across all
instances.
Within the present framework, the corresponding
algorithmic question concerns the preparation
of quantum distributions with useful
concentration on repair preimages and the
classical resources needed to transform
their samples into high-quality feasible
solutions, providing a new avenue
for analyzing complete quantum--classical procedures.

\subsection{Feasibility beyond exact permutation repair}
\label{subsec:beyond-hybrid-design}

The pushforward construction extends naturally
to problems requiring several successive
feasibility corrections. In such settings,
the raw sampling law is transformed through
a composition of problem-dependent classical
maps, each resolving a prescribed family
of constraint violations. Consider a sequence of repair operations
\[
    R_1,R_2,\ldots,R_k,
\]
whose composition maps the raw configuration
space into the desired feasible set,
\[
    \mathcal R
    :=
    R_k\circ\cdots\circ R_1.
\]
By the functoriality of pushforwards
established in
Proposition~\ref{prop:pushforward-properties},
\begin{equation}
    (R_k\circ\cdots\circ R_1)_\#p
    =
    (R_k)_\#
    \cdots
    (R_1)_\#p.
\end{equation}
Thus, the final repaired distribution can be
constructed through successive pushforwards
of the same underlying sampling law. For any such target set $G$ of feasible solutions requiring successive repairs, the complete repair procedure therefore defines a composite successful preimage in the original sampling space.

This formulation applies to constrained problems in which permutation correction, vehicle assignment, capacity restoration,
and route-contiguity correction are performed through successive
problem-dependent operations. When the individual maps are constructed to preserve previously satisfied constraints, the
composition provides a systematic route from encoded samples to globally feasible solutions. Each stage induces a further coarse-graining of the sampling law. This extends the repair-basin viewpoint
from a single feasibility map to a compositional architecture for hybrid constrained optimization.

\section{Conclusion}

We developed a distributional framework for constrained
hybrid quantum optimization in which deterministic repair
transports the raw quantum sampling law onto the feasible
set. The resulting pushforward distribution provides exact
expressions for repaired success probabilities and expected
objective values. Within the OFM kernel, we further
developed a prescription for constructing problem-dependent
repair maps from the constraint structure of the penalty
Hamiltonian, with Hungarian maximum-overlap repair
providing a concrete realization of the formalism for
permutation-constrained optimization problems.

Combining repair-basin aggregation with the shell-transfer
formalism from prior work\cite{onahseperating}, we established finite-shot
success guarantees and a basin-dependent coherent-depth
threshold. The analysis shows how the optimal-basin volume,
characterized by $\kappa_\star=n!|A_\star|/n^n$, modifies
the amplification requirement and reduces the corresponding
absolute-transfer depth threshold through the correction
$-\log\kappa_\star$. This establishes a quantitative
connection between classical repair geometry, coherent
quantum amplification, and the sampling resources required
for optimal-solution recovery
(\cref{thm:log-depth-threshold,subsec:discussion-pointwise-to-basin}).

% Although the maximizing mixer angle is known explicitly, lattice normalization and phase
% alignment provide an instance-dependent prescription
% for selecting the cost angle $\gamma$, which needs additional simulation to establish. These problem sizes remain beyond the computational
% reach of the quantum implementations and classical
% simulations available today. We hope that future
% advances in quantum hardware and classical simulation
% will enable direct verification of these finite-size
% sampling predictions.

The distributional mechanism is further characterized
by the exact factorization
$(R_\#p)(\pi^\star)=(\kappa_\star/n!)\chi_\star$,
which separates combinatorial repair-basin volume
from quantum probability concentration
(\cref{subsec:discussion-volume-concentration}).
Our numerical experiments provide a striking
illustration of the latter using single-layer
CE--QAOA and a coarse search over only $n+1$
candidate cost angles $\gamma$. Despite this
elementary parameter search, the repaired quantum
distributions exhibit systematically enhanced
concentration on low-cost tours relative to a
shot-matched encoded-uniform baseline subjected
to the same repair map. The repaired
cost CDFs and solution-level comparisons reveal
 probability concentration on useful
repair preimages. More strikingly, the contributor
histograms show individual infeasible configurations
whose quantum sampling counts exceed the expected
count for an entire average repair basin. These observations
demonstrate that pronounced instance-dependent
quantum concentration can already emerge from
very shallow circuits and inexpensive parameter
searches, complementing the modest combinatorial
basin enhancements considered in our finite-size
analysis (\cref{subsec:noiseless-analysis}).

For the largest noiseless instance, pushforward
repair recovered the optimum using approximately
$10^3$ shots, compared with the $10^6$ shots
reported in the original CE--QAOA simulations
without pushforward repair. Hardware experiments
on IBM Brisbane, using encodings of up to 100
logical variables, further recovered known optimal
tours from noisy measurement outcomes
(\cref{sec:experiment}) with $10,000$ shots. Beyond permutation
constraints, the compositional pushforward
construction provides a route to successive
feasibility corrections for more general
OFM-admissible optimization problems
(\cref{subsec:beyond-hybrid-design}).
These theoretical and experimental
results establish repair-basin geometry and
quantum probability concentration as complementary
resources for hybrid constrained optimization,
and provide a concrete pathway toward quantum
utility and quantum advantage through the joint
design of quantum sampling and classical
feasibility restoration.

% \sout{A significant example of resource enhancement arising from our proposal is the range of TSP problem sizes accessible (in principle) to single-layer CE--QAOA with polynomial sampling resources. Without pushforwards, at the transport-maximizing mixer angle $\beta^\star=\pi(n-1)/n$ and a shot budget of $n^3$, the pointwise absolute-transfer criterion for direct
% recovery of an optimal encoded bitstring is satisfied
% only through $n=13$ without going to additional alternating layers. Under our resource enhancement result (\cref{thm:log-depth-threshold}),
% deterministic repair extends the corresponding absolute-transfer criterion through $n=28$ at the average repair-basin scale ($f_n=1$),
% and through $n=30$ with a modest fivefold basin enhancement ($f_n=5$),
% as discussed in
% \cref{subsec:discussion-depth-phase-control} and
% \cref{app:basin-shot-depth-tradeoff}.
% Thus, exploiting the pushforward formalism more than doubles the problem-size threshold for absolute-transfer compensation while maintaining the same resource budget. This provides a quantitative explanation of the resource enhancement
% observed in our simulations. The analysis extends the absolute-transfer threshold to 30-city
% TSP instances with 900-qubit encodings, beyond the reach of currently available quantum hardware and classical quantum-circuit simulators. Future advances in both will enable direct investigation of these finite-size predictions under the phase-alignment conditions established in \cref{thm:log-depth-threshold}.

A concrete illustration of the repair-basin contribution is provided
by the finite-size absolute-transfer thresholds for single-layer
CE--QAOA. At the transport-maximizing mixer angle
$\beta_\star=\pi(n-1)/n$ and with a shot budget $N=n^3$, the
pointwise absolute-transfer criterion for a fixed optimal encoded
bitstring is satisfied through $n=13$. Deterministic repair extends
the corresponding criterion through $n=28$ at the average
repair-basin scale ($f_n=1$), and through $n=30$ for a fivefold
relative basin volume ($f_n=5$), as discussed in Sections~6.2
and~A.5. Thus, repair-basin aggregation more than doubles the
problem-size threshold for absolute-transfer compensation under
the same polynomial shot-budget rule.
These thresholds quantify the potential contribution of repair
geometry to the transfer budget. The corresponding fixed-confidence
sampling guarantees additionally require the phase-alignment
hypothesis of Theorem~8, the associated phase factor, and the
specified optimal-basin volume.  The $30$-city case corresponds to a $900$-qubit assignment encoding, providing a concrete target for further theoretical, numerical, and hardware investigation.

\section*{Data and code availability}

A minimal python implementation is available at
\href{https://doi.org/10.5281/zenodo.23694976}
{https://doi.org/10.5281/zenodo.21302533}. Additional data available upon request.

% The bit strings generated in the noiseless simulations and hardware experiments
% that support the numerical results of this work are available from the authors
% upon reasonable request.

\section*{Conflict of Interests}
All authors declare no competing interests.

\section*{Acknowledgements}
The authors gratefully acknowledge helpful discussions initiated at the International Workshop on Quantum Optimization held in Oslo, Norway in March 2026. S.H. acknowledges support from AFRL Contract No. FA8750-25-C-B0040.

\appendix
\section{Pushforward from deterministic repair}

\subsection{Structural properties of deterministic repair}
\label{app:exact-pushforward-properties}

Deterministic feasibility repair acts on sampling laws through a pushforward. Its structural properties follow from the repair map itself and apply to any probability law on the raw sample space. In noisy implementations, the pushforward remains well defined, while the raw distribution may lose encoded-subspace support or concentration on useful repair preimages.

Let $\mathcal X$ be a finite raw sample space, let $F\subseteq\mathcal X$ be the feasible set, and let $R:\mathcal X\to F$ be a deterministic repair map. For a probability law $p$ on $\mathcal X$, define
\[
    p_R:=R_\#p,
    \qquad
    p_R(A):=p(R^{-1}(A)),\quad A\subseteq F.
\]
In particular,
\[
    p_R(y)=\sum_{x:\,R(x)=y}p(x),\qquad y\in F.
\]

\begin{proposition}[Basic structural and pushforward properties of deterministic repair]
\label{prop:pushforward-properties}
Let $F\subseteq\mathcal X$ be finite and let $R:\mathcal X\to F$ satisfy $R(y)=y$ for every $y\in F$. For any probability law $p$ on $\mathcal X$, write $p_R=R_\#p$. Then:
\begin{enumerate}
    \item \emph{Positivity and normalization:}
    \[
       p_R(y)\geq0\quad(y\in F),\qquad
       \sum_{y\in F}p_R(y)=1.
    \]
    \item \emph{Basin aggregation:} with $B_y:=R^{-1}(\{y\})$,
    \[
        p_R(y)=\sum_{x\in B_y}p(x).
    \]
    \item \emph{Change of variables:} for any $C:F\to\mathbb R$ and $Z\sim p$,
    \[
      \mathbb E_{p_R}[C]
      =\sum_{y\in F}C(y)p_R(y)
      =\sum_{x\in\mathcal X}C(R(x))p(x)
      =\mathbb E_p[C(R(Z))].
    \]
    \item \emph{Functoriality:} for maps $R_1:\mathcal X\to F_1$ and $R_2:F_1\to F_2$,
    \[
      (R_2\circ R_1)_\#p
      =(R_2)_\#\bigl((R_1)_\#p\bigr).
    \]
    \item \emph{Retraction and idempotency:}
    \[
       R\circ R=R,\qquad R_\#(R_\#p)=R_\#p,
    \]
    where $R$ acts as the identity on $F$ in the second expression.
    \item \emph{Compositionally closed repair family:} let
    \[
      \mathfrak R_F
      :=\{R:\mathcal X\to F:R|_F=\operatorname{id}_F\}.
    \]
    For every $R,S\in\mathfrak R_F$,
    \[
      R\circ S=S,\qquad S\circ R=R.
    \]
    Thus $\mathfrak R_F$ is an idempotent semigroup under composition.
    \item \emph{Total-variation contraction:} for any probability laws $p,q$ on $\mathcal X$,
    \[
      \|R_\#p-R_\#q\|_{\mathrm{TV}}
      \leq\|p-q\|_{\mathrm{TV}}.
    \]
    \item \emph{Efficient samplewise realization:} if one repair takes time $T_R(n)$, then repairing $S$ independent samples requires
    \[
       T_{\mathrm{repair}}(S,n)=O\!\left(S\,T_R(n)\right).
    \]
    In particular, polynomial shot budgets and polynomial-time repair give a polynomial-time classical repair stage.
\end{enumerate}
\end{proposition}

\begin{proof}
Since $R(\mathcal X)\subseteq F$,
\[
   \sum_{y\in F}p_R(y)
   =p(R^{-1}(F))=p(\mathcal X)=1.
\]
Positivity and the basin formula follow immediately from the definition of $R_\#p$. Reordering the finite sums yields
\[
\begin{aligned}
   \sum_{y\in F}C(y)p_R(y)
   &=\sum_{y\in F}C(y)\sum_{x:R(x)=y}p(x)\\
   &=\sum_{x\in\mathcal X}C(R(x))p(x).
\end{aligned}
\]
For $A\subseteq F_2$, functoriality follows from
\[
\begin{aligned}
  ((R_2\circ R_1)_\#p)(A)
   &=p\bigl(R_1^{-1}(R_2^{-1}(A))\bigr)\\
   &=\bigl((R_2)_\#((R_1)_\#p)\bigr)(A).
\end{aligned}
\]
Because $R(x)\in F$ and $R$ fixes $F$, one has $R(R(x))=R(x)$. The corresponding pushforward identity then follows by functoriality. Likewise, for $R,S\in\mathfrak R_F$,
\[
   R(S(x))=S(x),\qquad S(R(x))=R(x),
\]
which proves closure and the stated composition identities. For total variation,
\[
\begin{aligned}
  \|R_\#p-R_\#q\|_{\mathrm{TV}}
  &=\sup_{A\subseteq F}\left|p(R^{-1}(A))-q(R^{-1}(A))\right|\\
  &\leq\sup_{B\subseteq\mathcal X}|p(B)-q(B)|
   =\|p-q\|_{\mathrm{TV}}.
\end{aligned}
\]
Finally, evaluating $R$ on each of $S$ samples gives the stated runtime bound.
\end{proof}

\subsection{Full-cube sampling via generic QAOA}
\label{subsec:generic-qaoa}

The pushforward formulation applies whenever measured configurations are mapped to feasible outputs before objective scoring. The sampled configuration space depends on the quantum ansatz. Generic QAOA samples the full Boolean cube, whereas the OFM-compatible CE--QAOA construction samples a prescribed one-hot product space.

For $N$ binary variables, generic QAOA~\cite{Farhi2014QAOA} prepares
\[
\ket{\psi_p(\vec\gamma,\vec\beta)}
=
\left[\prod_{\ell=1}^{p}
 e^{-i\beta_\ell\sum_{j=1}^N X_j}
 e^{-i\gamma_\ell H_C}\right]\ket{+}^{\otimes N}.
\]
The transverse-field mixer acts on $\Omega=\{0,1\}^N$, and the diagonal cost Hamiltonian generates problem-dependent phases. Techniques for selecting effective parameter schedules include linear or adiabatic ramps~\cite{%Lucas2014,
montanezbarrera2024universalqaoa}, parameter transfer across instances~\cite{Monta_ez_Barrera_2025T}, recursive variable elimination~\cite{Bravyi_2020}, and warm-start initialization~\cite{Egger2021Warm,carmo2025warmstarting}.

Let $R:\Omega\to F$ be any deterministic feasibility repair. The resulting law is
\[
   (R_\#p)(y)=\sum_{z\in\Omega:\,R(z)=y}p(z),
   \qquad y\in F.
\]
All structural pushforward identities apply unchanged. For CE--QAOA in the ideal OFM setting, the initial state and block-XY mixer restrict the raw law to $\mathcal X=[n]^m\subseteq\Omega$. This encoded structure also determines the basin-volume factors and shell-transfer expressions used in \cref{sec:subspace}. In noisy hardware implementations, the decomposition into encoded and leaked contributions is given in \cref{subsec:noise-leakage-basin-breakdown}.

\subsection{Tie-breaking and uniqueness issues}
\label{appendix:tie-breaking-uniqueness}

The deterministic repair map $R:\mathcal X\to F$ is single valued. For maximum-overlap assignment, this requires a fixed rule whenever several permutations have the same overlap with a raw word. In the TSP case study, Hungarian repair minimizes Hamming distance to the feasible permutation sector. A lexicographic convention, for example, defines
\[
  R(M)
  :=\min_{\mathrm{lex}}
    \arg\max_{P\in\mathcal P_n}\langle M,P\rangle.
\]
Any fixed total ordering gives a valid deterministic convention. The same rule must be applied to the quantum and encoded-uniform samples in a distributional comparison. For the encoded-uniform law $u_X(x)=n^{-n}$ on $\mathcal X=[n]^n$, every fixed deterministic rule gives
\[
  (R_\#u_X)(\pi)
  =\frac{|R^{-1}(\pi)|}{n^n}
  =\frac{\kappa_\pi}{n!},
  \qquad
  \kappa_\pi:=\frac{n!|R^{-1}(\pi)|}{n^n}.
\]
The repair basins partition $\mathcal X$, so
\[
   \frac{1}{n!}\sum_{\pi\in S_n}\kappa_\pi=1.
\]
The factor $\kappa_\star$ appearing in \cref{thm:log-depth-threshold} is the value corresponding to the selected permutation. For any sampling law $p$ on $\mathcal X$, define the basin concentration factor by
\[
   \chi_\pi
   :=\frac{(R_\#p)(\pi)}{(R_\#u_X)(\pi)}.
\]
The repaired probability then factorizes exactly as
\[
   (R_\#p)(\pi)=\frac{\kappa_\pi}{n!}\chi_\pi.
\]
For an optimal-cost set $G\subseteq S_n$, the uniform pushforward probability is similarly $|R^{-1}(G)|/n^n$.

\paragraph{A symmetry-controlled stochastic repair.}
\label{app:fair-repair}
Let $x=(x_1,\ldots,x_n)\in[n]^n$, define
\[
 D(x)=\{x_1,\ldots,x_n\},\qquad
 k(x)=|D(x)|,\qquad
 c_a(x)=|\{i:x_i=a\}|.
\]
Consider the stochastic maximum-overlap repair kernel
\begin{equation}
\label{eq:uniform-nearest-kernel}
K(\pi\mid x)
=
\frac{\mathbf1\!\left\{\sum_{i=1}^n\mathbf1\{x_i=\pi_i\}=k(x)\right\}}
     {(n-k(x))!\prod_{a\in D(x)}c_a(x)},
\qquad \pi\in S_n.
\end{equation}

\begin{proposition}[Uniform nearest repair]
\label{prop:uniform-nearest-repair}
For each $x$, the kernel $K$ selects uniformly among its maximum-overlap permutations and fixes feasible permutations. It can be sampled in $O(n)$ time. For the uniform encoded law $u_X$,
\begin{equation}
\label{eq:uniform-repair-control}
 (K_\#u_X)(\pi)=\frac1{n!},\qquad
 (K_\#u_X)(G)=\frac{|G|}{n!},\quad G\subseteq S_n.
\end{equation}
For any raw law $p$,
\[
 (K_\#p)(G)=\sum_{x\in[n]^n}p(x)K(G\mid x),
 \qquad K(G\mid x):=\sum_{\pi\in G}K(\pi\mid x).
\]
\end{proposition}

\begin{proof}
A permutation can match $x$ at most once for each distinct symbol in $D(x)$. Choose one of the $c_a(x)$ occurrences for every such symbol $a$, then assign the $n-k(x)$ missing symbols uniformly to the remaining positions. This produces exactly
\[
  (n-k(x))!\prod_{a\in D(x)}c_a(x)
\]
maximum-overlap permutations with equal probability. The selection requires $O(n)$ operations. If $x\in S_n$, then $k(x)=n$ and the unique output is $x$.

For any relabeling $g\in S_n$, the nearest-permutation sets of $x$ and $g\cdot x$ are in bijection, so
\[
 K(g\cdot\pi\mid g\cdot x)=K(\pi\mid x).
\]
The uniform word law is invariant under relabeling, and relabeling acts transitively on $S_n$. Hence $K_\#u_X$ is uniform on $S_n$. Summation over $G$ gives the remaining identity.
\end{proof}

For the symmetry-controlled stochastic repair $K$ of
Proposition~10, if the undirected optimum is unique up to rotation and reversal,
and all such encodings are retained, then $|G_\star|=2n$. For a specified good set $G$, the quantity $K(G\mid X)$ with $X\sim p$ is an unbiased estimator of $(K_\#p)(G)$. For an explicitly listed $G$, it can be evaluated in $O(n|G|)$ time. It is the conditional expectation of the hit indicator obtained from a random tied repair and therefore has no greater variance. If an undirected optimal tour is unique up to rotations and reversal, its $2n$ permutation encodings have encoded-uniform baseline probability $2n/n!$ under $K$.

For $n\geq3$, no deterministic retraction $[n]^n\to S_n$ is equivariant under all city relabelings. A constant word is fixed by a transposition of two unused city labels, whereas no permutation is fixed by that relabeling. Hence the uniformity in \eqref{eq:uniform-repair-control} is a property of the stochastic kernel and does not imply equal basin cardinalities for a fixed deterministic tie-breaking rule.

\subsection{Problem-dependent statistics}
\label{app:problem-stat}

% \sh{The procedures described below are candidate repair strategies.
% Their use as efficient total repairs requires separate guarantees
% of termination, feasibility, and computational cost. When several
% repairs are composed, later stages must also preserve the
% constraints enforced by earlier stages.}

\subsubsection{Contingency-table and transportation constraints}

Fixed-margin contingency tables are specified by prescribed row sums, column sums, and higher-order marginals~\cite{DiaconisSturmfels1998,DeLoeraOnn2004,SlavkovicZhuPetrovic2015}. Related integer-programming constructions arise through multiway transportation formulations~\cite{DeLoeraOnn2006}. These marginals define the constraint system and enter the penalty Hamiltonian through their residuals. A suitable structural statistic records the observed marginals and their deviations from the required values.

Such residual information identifies entries carrying excess mass and entries with corresponding deficits. Local exchanges, cycle corrections, and flow-balancing operations can then restore prescribed margins. The resulting pushforward law aggregates the probabilities of all measured tables that repair to the same feasible table or transportation plan.

\subsubsection{Coloring and constraint-satisfaction problems}

Graph-coloring constraints are described through assigned colors, monochromatic edges, vertex conflict degrees, color populations, and overlaps with reference colorings. Related overlap-matrix statistics appear in random coloring and constraint-satisfaction problems~\cite{AchlioptasCojaOghlan2008,CojaOghlanVilenchik2016}. The constituent edge-conflict and assignment residuals identify the vertices and colors associated with each violation.

A problem-dependent repair can recolor high-conflict vertices, preserve valid neighborhoods, and select replacement colors using the local conflict profile. The repaired objective, correction count, and output coloring may then be analyzed as functions of those statistics. The pushforward law aggregates the raw probability of all assignments repairing to the same feasible coloring.

\subsubsection{Routing and assignment formulations}

Routing and assignment formulations contain multidimensional incidence statistics. In a position-based TSP encoding, the relevant quantities include city-visit populations, repeated and missing cities, and assignment marginals. Arc-based formulations additionally involve incoming and outgoing degree residuals, connectivity conditions, and subtour structure. The Miller--Tucker--Zemlin formulation introduces ordering variables and residuals to eliminate subtours~\cite{MillerTuckerZemlin1960}, while many-visits TSP imposes prescribed visit multiplicities~\cite{CosmadakisPapadimitriou1984}.

The corresponding penalty residuals identify repeated cities, omitted cities, degree violations, disconnected subtours, etc. Repair operations can replace repeated cities, correct degrees, reconnect routes, and insert omitted visits. Assignment and matching problems admit an analogous description through row and column populations, unmatched elements, collisions, and overlap with feasible assignments. Maximum-overlap matching and minimum-distance projection provide natural repair primitives.

Vehicle-routing problems introduce customer coverage, vehicle assignment, route load, capacity, depot incidence, connectivity, and route-count statistics. Their residuals can guide customer transfers, route splitting, component reconnection, insertion of unserved customers, and elimination of duplicate visits. The corresponding pushforward distribution identifies the raw configurations contributing to feasible routes and supports the analysis of objective changes, and concentration on high-quality repaired outputs.

\subsubsection{Scheduling, allocation, and packing formulations}

Scheduling constraints involve job multiplicities, machine assignments, machine loads, time-slot populations, precedence relations, processing orders, and resource consumption. Configuration and $N$-fold integer-programming formulations organize these through job-type, machine-load, and time-slot aggregates~\cite{MnichWiese2015,KnopKoutecky2018}. Penalty residuals record unassigned jobs, duplicate assignments, machine overloads, time conflicts, precedence violations, and resource excesses.

These statistics can guide the assignment of missing jobs, removal of duplicates, transfers between machines, resolution of time conflicts, and restoration of precedence. Allocation, generalized-assignment, multiple-knapsack, and packing formulations admit a similar construction using assignment populations, resource-usage vectors, capacity residuals, and unallocated-item counts. The repair map then transports raw probability to feasible allocations according to the full violation profile.

\subsection{Basin enhancement, shot complexity, and delayed depth crossover}
\label{app:basin-shot-depth-tradeoff}

The finite-depth compensation threshold depends on the volume of the optimal repair preimage and the available sampling budget. Here we evaluate this relation explicitly for a single CE--QAOA layer, as used in the noiseless and hardware experiments. For $n\geq4$ and $\mathcal X=[n]^n$, fix an optimal permutation $\pi^\star$ and its basin $A_\star=R^{-1}(\pi^\star)$. Write
\begin{equation}
\label{eq:basin-enhancement-factor}
  |A_\star|=f_n\frac{n^n}{n!},
  \qquad f_n:=\kappa_\star=\frac{n!|A_\star|}{n^n}.
\end{equation}
Thus $f_n=1$ is the average-basin reference, while $f_n>1$ describes an optimal basin larger than that average. The definition uses the same fixed deterministic repair map as the numerical comparisons.  Under the phase-alignment hypothesis of Corollary~\ref{cor:uniform-phase-aligned-repair}, the depth-one repaired-success bound specializes to
\begin{equation}
\label{eq:p1-basin-enhanced-success}
 P^{\mathrm{rep}}_{\mathrm{CE},1}(\pi^\star)
 \geq
 \frac{f_n}{n!}\left(3-\frac4n\right)^{2n}
 \cos^2\!\left(\frac{\Theta_1}{2}\right).
\end{equation}
The basin factor contributes multiplicatively to the depth-one absolute-transfer term.

The absolute-transfer factor reaches the basin-normalization scale at
\begin{equation}
\label{eq:p1-constant-shot-condition}
  f_n\left(3-\frac4n\right)^{2n}\geq n!.
\end{equation}
Equivalently, the required relative basin volume is
\begin{equation}
\label{eq:required-basin-enhancement}
  f_{\mathrm{req}}(n)
  :=\frac{n!}{(3-4/n)^{2n}}.
\end{equation}
For selected finite sizes,
\begin{equation}
\label{eq:basin-enhancement-numerics}
\begin{array}{c|rrrrrrrr}
 n&18&19&20&21&22&23&24&25\\\hline
 f_{\mathrm{req}}(n)&
 0.681&1.430&3.161&7.341&17.870&45.495&120.898&334.756
\end{array}
\end{equation}
At the average-basin scale $f_n=1$, the depth-one absolute-transfer compensation criterion holds through $n=18$. Increasing $f_n$ to $2$, $5$, or $10$ extends this crossover through $n=19$, $20$, or $21$, respectively. These values locate the crossover of the absolute-transfer budget at the maximizing angle.

\paragraph{Polynomial-shot compensation.}
Consider $N(n)=n^k$ independent samples. The depth-one absolute-transfer budget, including the sampling allowance, is
\begin{equation}
\label{eq:p1-transport-budget-appendix}
 \mathcal B_1(n;f_n,k)
 :=\frac{f_n n^k}{n!}
 \left(3-\frac4n\right)^{2n}.
\end{equation}
The finite-size threshold plotted in \cref{fig:p1-pushforward-budget} is $\mathcal B_1(n;f_n,k)\geq1$. In terms of the required relative basin volume, this becomes
\[
    f_n\geq\frac{n!}{n^k(3-4/n)^{2n}}.
\]
This criterion separates the effect of repair-basin volume, coherent absolute transport, and polynomial sampling resources.

For completeness, the corresponding phase-aligned probability bound gives an explicit confidence relation. Define $c_1:=\cos^2(\Theta_1/2)$. If the common-arc hypothesis holds on $A_\star$, then $N$ independent shots fail to recover $\pi^\star$ with probability at most
\[
\begin{aligned}
  \Pr(\mathrm{failure})
  &=\left(1-P^{\mathrm{rep}}_{\mathrm{CE},1}(\pi^\star)\right)^N \leq\exp\!\left[-\frac{N f_n}{n!}
            \left(3-\frac4n\right)^{2n}c_1\right].
\end{aligned}
\]
Consequently, for $N=n^k$ and target failure probability $\epsilon$, a sufficient condition under that hypothesis is
\begin{equation}
\label{eq:p1-polynomial-shot-condition}
 f_n n^k\left(3-\frac4n\right)^{2n}
 \cos^2\!\left(\frac{\Theta_1}{2}\right)
 \geq n!\log\!\left(\frac1\epsilon\right).
\end{equation}
The transport budget $\mathcal B_1$ records the geometric and mixer contributions; the phase factor and desired confidence enter the corresponding success guarantee through \eqref{eq:p1-polynomial-shot-condition}.

The largest sizes satisfying $\mathcal B_1\geq1$ for representative relative basin volumes and polynomial shot budgets are
\begin{equation}
\label{eq:shot-basin-crossover-table}
\begin{array}{c|rrrr}
\text{shot budget}& f_n=1&f_n=2&f_n=5&f_n=10\\\hline
N=1    &18&19&20&21\\
N=n    &22&23&23&24\\
N=n^2  &25&26&27&27\\
N=n^3  &28&29&30&30\\
N=n^5  &34&35&36&36\\
N=n^{10}&48&49&49&50
\end{array}
\end{equation}
These evaluations establish a finite-size tradeoff between optimal-basin volume and sampling allowance at fixed coherent depth $p=1$ and fixed maximizing mixer angle $\beta=\beta^\star(n)$. Larger basins and larger polynomial shot budgets extend the range satisfying the depth-one absolute-transfer criterion. The corresponding certified repaired-success probability is determined by the full bound \eqref{eq:p1-basin-enhanced-success}.

\bibliographystyle{apsrev4-2}
\bibliography{ref}
\end{document}